\documentclass[usenatbib]{mn2e} \usepackage{psfig,longtable}

\title[Estimating the masses of extra-solar planets] {Estimating the masses of extra-solar planets}

\author[C.\,A.\ Watson,  S.\,P.\ Littlefair, A.\ Collier Cameron,
V.\,S.\ Dhillon and E.\,K.\ Simpson]
{C.\,A.\ Watson,$^1$\thanks{E-mail: c.a.watson@qub.ac.uk}
S.\,P.\ Littlefair,$^2$ A.\ Collier Cameron,$^3$
V.\,S.\ Dhillon,$^2$ and E.\,K.\ Simpson$^1$ \\
$^1$ Astrophysics Research Centre, School of Mathematics \& Physics, Queen's University,
University Road, Belfast BT7 1NN, UK \\
$^2$ Department of Physics and Astronomy, University of Sheffield,
Sheffield S3 7RH, UK \\
$^3$ School of Physics and Astronomy, University of St Andrews,
North Haugh, St Andrews, Fife KY19 9SS, UK \\}

\date{\center{\Large Accepted for publication in the Monthly Notices
    of the Royal Astronomical Society \\
\vspace{.5cm} \today}}

\begin{document}
\maketitle
 
\begin{abstract}

All extra-solar planet masses that have been derived spectroscopically
are lower limits since the inclination of the orbit to our
line-of-sight is unknown except for transiting systems.  In theory,
however, it is possible to determine the inclination angle, $i$,
between the rotation axis of a star and an observer's line-of-sight
from measurements of the projected equatorial velocity ($v \sin i$),
the stellar rotation period ($P_{rot}$) and the stellar radius
($R_*$).  For stars which host planetary systems this allows the
removal of the $\sin i$ dependency of extra-solar planet masses
derived from spectroscopic observations under the assumption that the
planetary orbits lie perpendicular to the stellar rotation axis.

We have carried out an extensive literature search and present a
catalogue of $v \sin i$, $P_{rot}$, and $R_*$ estimates for stars
hosting extra-solar planets. In addition, we have used Hipparcos
parallaxes and the Barnes-Evans relationship to further supplement the
$R_*$ estimates obtained from the literature. Using this catalogue, we
have obtained $\sin i$ estimates using a Markov-chain Monte Carlo
analysis. This technique allows proper 1-$\sigma$ two-tailed
confidence limits to be placed on the derived $\sin i$'s along with
the transit probability for each planet to be determined.

While we find that a small proportion of systems yield $\sin i$'s
significantly greater than 1, most likely due to poor $P_{rot}$
estimations, the large majority are acceptable. We are further
encouraged by the cases where we have data on transiting systems, as
the technique indicates inclinations of $\sim$90$^{\circ}$ and high
transit probabilities. In total, we are able to estimate the true
masses of 133 extra-solar planets. Of these 133 extra-solar planets,
only 6 have revised masses that place them above the 13 $M_J$
deuterium burning limit; 4 of those 6 extra-solar planet candidates
were already suspected to lie above the deuterium burning limit before
correcting their masses for the $\sin i$ dependency. Our work reveals
a population of high-mass extra-solar planets with low eccentricities
and we speculate that these extra-solar planets may represent the
signature of different planetary formation mechanisms at work.
Finally, we discuss future observations that should improve the
robustness of this technique.

\end{abstract}


\section{Introduction}
\label{sec:intro}

Over 16 years ago the first planets to be detected outside of our
solar system were discovered around the millisecond pulsar PSR1257+12
(\citealt{wolszcan92}). Within 3 years, \cite{mayor95} announced the
first planet orbiting around a main-sequence star, 51 Peg b. Since
then, extra-solar planet candidates have been discovered at a
phenomenal rate. At the time of writing, 453 extra-solar planet
candidates have now been identified through a variety of techniques
including radial velocity studies, transits, microlensing events,
stellar pulsations as well as pulsar timing.

By far the most extra-solar planets have been discovered by observing
the small Doppler wobble of the host star. This technique, however,
only returns a minimum mass $M \sin i$ (where $M$ is the mass of the
planet, and $i$ is the inclination of the normal to the planetary
orbital plane to the observer's line-of-sight), which is a firm lower
limit to the true planetary mass. Indeed, the inclination (and hence
true planetary mass) can only be determined accurately for those
planets which transit their host star. With only $\sim$ 70 transiting
planets known, this leaves the vast majority of planets with only
lower-limits placed on their masses. Improving the mass determinations
of these planets has obvious benefits for planet formation modeling
and for studying the planet mass distribution.

In this paper we present a method for estimating the orbital
inclinations and hence true masses of non-transiting extra-solar
planets. We then apply this method to the extra-solar planet systems
for which there is sufficient data available, and investigate the
impact that the corrected masses have on our knowledge of extra-solar
planet properties. Finally, we conclude with a look at the  improved
measurements that should be made to make this technique more robust.

\section{Estimating the orbital inclinations of extra-solar planets}
\label{sec:technique}

It is possible to determine the inclination angle, $i$, between the
rotation axis of the extra-solar planet host star and the observer's
line-of-sight. By combining measurements of the star's projected
equatorial velocity ($v \sin i$), the stellar rotation period
($P_{rot}$) and the stellar radius ($R_*$) one can determine $\sin i$
from

\begin{equation}
\sin i = \frac{P_{rot} \times v \sin i}{2 \pi R_*}.
\label{eqn:sini}
\end{equation}
Indeed, this method has previously been applied by \cite{gonzalez98}
to 7 exoplanet host stars, as well as by \cite{cameron97} to determine the
inclination of the rotation axis of the extensively Doppler-imaged young star
AB Dor, for example. Equation~\ref{eqn:sini} can then be used to lift the
$\sin i$ degeneracy in calculating extra-solar planet masses using
spectroscopic observations if it is assumed that the planetary orbits
lie perpendicular to the host star's rotation axis. 
Certainly, this condition holds true for our solar system, which has an
angle between
the plane of the ecliptic and the solar equator of around 7$^{\circ}$
(\citealt{beck05}). The degree of alignment between the stellar spin
axis and the planetary orbit can also be measured for transiting
extra-solar planets using the Rossiter-McLaughlin effect
(e.g.~\citealt{gaudi07}). So far this has been carried out for 26
planet systems (see \citealt{winn05}; \citealt{winn07};
\citealt{wolf07}; \citealt{narita07}; \citealt{johnson08};
\citealt{cochran08}; \citealt{hebrard08}; \citealt{bouchy08};
\citealt{winn08}; \citealt{johnson09}; \citealt{winn09};
\citealt{narita09}; \citealt{pont09}; \citealt{triaud09};
\citealt{gillon09}; \citealt{narita10}; \citealt{anderson10};
\citealt{jenkins10}); \citealt{simpson10}; \citealt{queloz10};
\citealt{triaud10}).

Of these 26 systems, 7 appear to have appreciable mis-alignment
angles.  These are HD 80606b, XO-3b, HAT-P-7b, WASP-2b, WASP-8b,
WASP-14b, WASP-15b and WASP-17b. However, \cite{hebrard08} suggest
that the spin-orbit misalignment measured for XO-3 may be due to a
systematic error as a result of the high airmass at which their
observations were carried out. We should also note that at first the
spin-orbit misalignment of HD 17156 was measured to be 62$^{\circ} \pm
25^{\circ}$ by \cite{narita08}, though the more recent work by
\cite{cochran08} concluded that the planetary orbital axis is, in
fact, very well aligned with the stellar rotation axis. \cite{pont09}
have reported a $\sim50^{\circ}$ misalignment in HD 80606. This system
is a binary, and the misalignment may well arise through the action of
the Kozai mechanism (e.g. \citealt{takeda05}; \citealt{malmberg07}).
HD 80606b also exhibits a large orbital eccentricity, no doubt as a
result of the Kozai interactions. In addition, WASP-8b is part of a
triple system (\citealt{queloz10}) and therefore its mis-alignment
angle is also most likely due to the Kozai mechanism. This leaves 4
planetary systems with confirmed mis-alignment angles for which no
stellar companion is yet known. Whether the Kozai mechanism is a
dominant process affecting the orbital evolution of exoplanets in
non-binary systems is yet to be seen, but obviously some caution must
be applied when assuming spin-orbit alignment. For now, however, we
will work on the premise that this assumption is a reasonable one for
single stars.

In order to measure the orbital inclination of extra-solar planets, we
can see from equation~\ref{eqn:sini} that we require just 3
quantities, $v \sin i, R_*$ and $P_{rot}$. The projected stellar
equatorial rotation-velocity, $v \sin i$, can be measured using high
resolution spectroscopy.  While the stars targeted by extra-solar
planet hunts are generally slowly rotating (in order to avoid spurious
radial velocities introduced by magnetic activity generated in rapidly
rotating stars), the spectrographs used for hunting extra-solar
  planets are high-resolution instruments. Thus most extra-solar
planet host stars have their line-broadening measured. One possible
caveat with these measurements is that the stellar rotation may no
longer be considered the sole line-broadening mechanism and other
mechanisms, such as turbulence (see Section~\ref{sec:disc} for a
discussion), may have to be taken into account.

The radii of the extra-solar planet host stars can be estimated in a
variety of ways. While some stars may have their radii measured
directly via interferometry, lunar occultations or transits/eclipses
(e.g.  \citealt{fracassini01}), the majority are estimated using
indirect methods. The most common method is to combine stellar
luminosities derived from bolometric corrections and {\em Hipparcos}
parallaxes with effective temperatures (determined from spectral
synthesis modeling) to determine the stellar radii. Indeed,
\cite{fischer05} have done exactly this for a large number of
extra-solar planet host stars, and quote a  median error on the radii
of $\sim$3 per cent.

In addition to the published values of the stellar radii, we have also
used the Barnes-Evans technique to estimate the angular diameters of
the extra-solar planet host stars. We have used the (V-K) colour --
angular diameter relation of \cite{fouque97}, who established the
following empirical surface brightness ($F_v$) -- colour relationship:

\begin{equation}
F_v = 3.947 - 0.131 \left( V-K \right).
\label{eqn:fv}
\end{equation}

\noindent When combined with the absolute visual magnitude, $M_v$, the
surface brightness parameter $F_v$ calculated in equation~\ref{eqn:fv}
can be used to determine the radius of the star, in solar radii, using
equation 2 of \cite{beuermann99}:

\begin{equation}
R_* = 10^ { 0.2 \times \left[ 42.368 - \left( 10 \times F_v \right) -
    M_v \right]}.
\end{equation}

\noindent Thus, only the $M_v$ of the host star is required, which can
be calculated from the $V-$band magnitude and parallax measurements
from Hipparcos. We have also taken into account extinction using the
reddening law from \cite{fouque97}:

\begin{equation}
E(V-K) = 0.88A_v,
\end{equation}

\noindent and the absorption law from \cite{diBenedetto87}:

\begin{equation}
A_v = 0.14 \times \frac{1- \exp \left( -10 \times d \times | \sin b |
  \right)} {|\sin b|},
\end{equation}

\noindent where $A_v$ is the visual absorption coefficient, $E(V-K)$
the $V-K$ colour extinction, $d$ the distance to the star in kpc and
$b$ the galactic latitude. We note that the stellar radii and
associated error bars we derive from the Barnes-Evans technique are in
excellent agreement with the published stellar radii for extra-solar
planet host stars showing an rms scatter of 6.7 per cent. This scatter
is largely Gaussian in nature, except for a number of notable
outliers. Indeed, on close inspection we find that out of the 373
individual stellar radii measurements presented in this work, 11
disagree with the Barnes-Evans derived radii by $3-sigma$ or more.
Statistically we would not expect more than 1 or 2 measurements to lie
beyond 3-$\sigma$. On closer inspection, apart from HD 41004A, all of
the outliers (HD 6434, HD 33283 (2 discrepant measurements), HD 33564,
HD 82943, HD 89744, HD 128311, HD 145675, HD 186427 and HD 216437)
have other radii measurements which agree well with the Barnes-Evans
derived radius. We can only surmise that these discrepant points are,
therefore, due to systematics.

This leaves one final quantity, the rotation period of the star,
$P_{rot}$, to be determined. Unfortunately, for the reasons stated
earlier, the majority of stars targeted in extra-solar planet hunts
are not highly active stars. Therefore, their rotation periods
generally cannot be measured by tracking of large, cool starspots on
their surfaces, for example. They are often, however, sufficiently
active to show Ca {\sc ii} H and K emission in their
spectra. \cite{noyes84} derived the ratio, $R'_{HK}$, of Ca {\sc ii} H
and K chromospheric emission to the total bolometric emission for a
number of stars whose rotation periods were known from variability in
their light curves. They found that, as expected from stellar dynamo
theory, the mean level of Ca {\sc ii} H and K emission is correlated
with rotation period.  In addition, the emission also depends on the
spectral type (probably due to convective zone depth). \cite{noyes84}
were then able to determine the following rotation period -- activity
relationship for main-sequence stars,

\begin{equation}
\log \left( P_{rot}/\tau \right) = 0.324 - 0.400y - 0.283y^2 -
1.325y^3,
\label{eqn:relation}
\end{equation}

\noindent where $y=\log (10^5 R'_{HK})$. The value for the convective
turnover time, $\tau$, can be obtained from the empirical function,

\begin{equation}
\log \tau = \left\{
\begin{array}{ll}
1.362 - 0.166x + 0.025x^2 - 5.323x^3 & :x > 0 \\ 1.362 -0.14x & :x < 0
\\ \end{array} \right.
\end{equation}

\noindent where $x=1-(B-V)$. Thus the stellar rotation period can be
determined from equation~\ref{eqn:relation} if $R'_{HK}$ and the $B-V$
colours are known.

We are in the fortunate position that many of the extrasolar planet
hosts have published $R'_{HK}$ values, since investigators generally
wish to show that the host stars exhibit low-level magnetic activity
and hence discard activity as the cause of radial velocity
variations. Furthermore, most extra-solar planet hosts are bright
stars, of which several have been observed by long-term surveys such
as the Mount Wilson H-K survey that started in the mid-1960's
(\citealt{wilson78}). Since the level of Ca {\sc ii} H and K emission
may vary with time due to, for example, solar-like activity cycles or
rotation of magnetic regions, $R'_{HK}$ measurements need to be
averaged over a suitably long ($\sim$decade) baseline. Given a
suitable span of observations, \cite{noyes84} found that they could
predict the rotation periods of stars with a reasonably high accuracy.
Obviously, for stars where only a few $R'_{HK}$ observations have been
made, the error on the rotation period may be much higher due to
intrinsic variability in the Ca {\sc ii} H and K emission. This is
discussed in section~\ref{sec:rhk_data}

\section{Application to known extrasolar planets}
\label{sec:appl}

In order to calculate the $\sin i$'s of the extra-solar planet hosts,
we have conducted an intensive literature and database search to
determine the 3 quantities $v \sin i, R_*$ and $\log R'_{HK}$. The
values we have found are presented in Table~\ref{tab:1}. Extra-solar
planet host stars for which we could not find estimates of all 3
quantities ($v \sin i, R_*$ and $\log R'_{HK}$) are not presented in
this table. Where identifiable, we have attempted to remove any
duplicate measurements. For example, many of the $v \sin i$
measurements taken from the NASA Stellar Archive and Exoplanet
Database (NStED -- see {\tt http://nsted.ipac.caltech.edu/}) were
found to be rounded values from \cite{fischer05} and have therefore
not been included in Table~\ref{tab:1} in these cases.

The values in Table~\ref{tab:1} have then been used to determine $v
\sin i$, $R_*$ and $P_{rot}$ for each star in our sample to obtain
$\sin i$ via equation~\ref{eqn:sini} as follows.  We have taken a
weighted mean for the final values of $v \sin i$ and $R_*$ (the latter
includes our radius estimate derived from the Barnes-Evans
technique). Where no error was quoted for a value of $v \sin i$ we
have taken it to be 1.0 km s$^{-1}$, which is twice the typical error
assumed on $v \sin i$ measurements (see the catalogue of
\citealt{fischer05}, for example). Regarding radii with no associated
error estimate, we have taken the error to be 10 or 20 per cent of the
absolute value. We have chosen 10 per cent when the only radius
measurement/s available for a particular star do not indicate
uncertainties. Where there is more than one radius estimate for a
star, of which one or more do not include error bars, then we have
assumed the error bar to be either 10 or 20 per cent. We chose whether
to adopt a 10 or 20 per cent uncertainty such that radii estimates
with associated error bars were given a higher weighting than those
without formal error bars in the final weighted mean.

\subsection{Adopted $\log R'_{HK}$ values and errors}
\label{sec:rhk_data}

The adopted values and error estimates for the $\log R'_{HK}$
measurements require special mention. A comprehensive literature
search has been conducted and for each $\log R'_{HK}$ measurement
reported in Table~\ref{tab:1} we have determined, where possible, the
number of observations and period span over which they were carried
out.  This detailed information is summarised in Table~\ref{tab:rhk}.
Where details of the $\log R'_{HK}$ measurements are either not
present or are ambiguous, we have assumed that they are from a single
observation and have flagged them as {\em `individual?'}. Where
available we have also quoted any reported variations or error
estimations in either the S-index (see \citealt{wright04} for the
definition of S-index, but note that their equation 10 is in error and
the left-hand side should read $\log C_{cf}(B-V) = \ldots$) or $\log
R'_{HK}$ measurement. These reported errors  should be treated with
caution since in many cases they only represent the measurement
accuracy and do not sample variations in the Ca H \& K emission over
the course of the stellar rotation and/or activity cycle.

After establishing how well monitored each star was, they were then
assigned a grade of P (Poor), O (O.K.), G (Good) or E (Excellent). A
grade of 'poor' was assigned to stars with only a few individual $\log
R'_{HK}$ measurements which would not be sufficient to sample the
variation of chromospheric emission throughout a stellar rotation. 'O'
was assigned to stars with a few observations spaced over several
months where the stellar rotation was probably adequately sampled, but
not the activity cycle. A grade of 'good' was assigned to stars with
more than 2 years worth of observations where  the stellar rotation
would be well sampled, but probably only a portion of any activity
cycle present had been covered. Finally, a grade of 'excellent' was
assigned to objects with over a decade of $\log R'_{HK}$ measurements
available which covered any likely activity cycle.

\cite{vaughan81} present a study of chromospheric Ca H \& K variations
as a function of stellar rotation for 46 lower main sequence field
stars. Their results show that, on average, rotation causes the
modulation of the S-index (and therefore also the $\log R'_{HK}$
measurements) by 7.3 per cent for F-stars, 9.4 per cent for G-stars,
and 13 per cent for K-stars. We refer to these values as the {\em
  average rotationally modulated variations} or ARMV. In addition,
\cite{vaughan81} show that modulations due to activity cycles are
typically twice that caused by rotation. We have used this to assign
general error bars on the $\log R'_{HK}$ values for our stars
dependent upon their spectral type and assigned grade (P, O, G, or E)
as follows:

\begin{itemize}
\item Grade P: 2.0 $\times$ the ARMV,
\item Grade O: 1.5 $\times$ the ARMV,
\item Grade G: 1.0 $\times$ the ARMV,
\item Grade E: 0.5 $\times$ the ARMV.
\end{itemize}

\noindent Thus, stars with only a few individual observations are
assigned an error that would cover the entire range in Ca H \& K
variations seen over a typical activity cycle. The error bars assigned
to the other categories are somewhat ad-hoc, but signify an
improvement in the reliability of the average $\log R'_{HK}$ as the
sampling of the activity cycle is improved. Given the amalgamation of
sources for the $\log R'_{HK}$ observations, we feel this is as robust
an error treatment that the data can be given in most cases. For
objects with several independent $\log R'_{HK}$ measurements, this
error assignment generally covers the observed variations well. In the
few cases where they do not, we have expanded the error bar to cover
the observed $\log R'_{HK}$ variations appropriately. Finally, for
objects whose activity cycles have been well monitored and for which
we can define a maximum variation across the cycle, we have taken
these limits as representing the 3-$\sigma$ variation on the average
$\log R'_{HK}$ value. (For example, if a well sampled star has a mean
$\log R'_{HK}$ = -4.9 but varies from -4.8 -- -5.0, we assigned a
1-$\sigma$ error = 0.1/3).

Where two or more $\log R'_{HK}$ measurements are available we have
taken a weighted mean of their values. The weightings are based on
either how many observations have been taken, or the time span over
which the observations were taken, depending on what information
exists.  We have then calculated the stellar rotational period using
the \cite{noyes84} relationship and $B-V$ values from the NStED
database. The rotation periods and the associated error bars we have
calculated are presented in Tables~\ref{tab:2} \&~\ref{tab:3} and can
be compared to the rotation periods obtained in the literature shown
in Table~\ref{tab:1}. Note, however, that we found several cases where
authors have clearly calculated the rotation period from $\log
R'_{HK}$ incorrectly (see Appendices A and B).

\section{Markov-chain Monte Carlo analysis}
\label{sec:mcmc}

Equation~\ref{eqn:sini}~can be thought of as a naive estimator of
$\sin i$. By simply inputting the derived values for $v \sin i$, $R_*$
and $P_{rot}$ for each host star (as discussed earlier) it is possible
to obtain an unconstrained distribution of $\sin i$ values
(i.e. values of $\sin i > 1$ are possible). Due to uncertainties in $v
\sin i$, $R_*$ and $P_{rot}$, this naive estimator will, however,
occasionally yield unphysical $\sin i$ values greater than 1.
Table~\ref{tab:2} lists all the exoplanet host stars which yield a
$\sin i > 1$ as calculated from equation~\ref{eqn:sini}, along with
their formal error bars.

For the purposes of this paper, however, we wished to carry out a
Markov-chain Monte Carlo (hereafter MCMC) analysis on the extra-solar
planet host stars. MCMC has the major advantage over simply using our
naive estimator (equation~\ref{eqn:sini}) in that, not only does it
provide a means of optimizing the fit of a model to data, but it also
explores the joint posterior probability distribution of the fitted
parameters.  This means that proper 1-$\sigma$ two-tailed confidence
limits can be placed on the derived $\sin i$'s, as well as allowing
the probability of a transit being observed to be calculated from
purely spectroscopic data. MCMC has been used in several areas of
astronomy, and instead of outlining in detail its operation here, we
refer the readers to \cite{tegmark04}, \cite{ford06} and
\cite{gregory07}, who have applied MCMC to various astronomical
problems including deriving cosmological parameters from the cosmic
microwave background, and deriving physical parameters of extra-solar
planet systems. In particular, our version of MCMC is modified from
the code used by \cite{cameron07} to identify extra-solar planet
transit candidates.

Naturally, values of $\sin i > 1$ are unphysical, and the MCMC rejects
those combinations of parameters that result in $\sin i > 1$.  If,
however, we imagine the hypothetical case where we have a population
of transiting extra-solar planets all with $\sin i = 1$ then, due to
measurement errors, on average half of these systems would yield $\sin
i > 1$  from equation~\ref{eqn:sini}. Obviously we would not want to
reject these systems on this basis, since they do not contradict our
null hypothesis that the measurements are free from systematic
errors. One particular example of this is HD209458, which is a known
transiting planet and yields $\sin i = 1.096 \pm 0.108 $ from our
naive estimator, equation~\ref{eqn:sini} (see Table~\ref{tab:2}). We
do, however, want to reject those systems where it is likely that
there are systematic errors in their $R_*, P_{rot}$ and $v \sin i$
measurements leading to $\sin i > 1$. We have, therefore, included all
systems from Table~\ref{tab:2} which are within 1-$\sigma$ of $\sin i
= 1$ in our MCMC analysis and have error bars $<$ 0.5.

For the purposes of this paper, we feed the MCMC with the measured
values of $R_*$, $P_{rot}$, $v \sin i$ and their associated error
bars, $\sigma_R$, $\sigma_P$, $\sigma_v$, respectively. We assume that
the stellar inclinations are randomly distributed and hence follow a
uniform distribution with $0 < x < 1$, where $x$ = $\cos i$. For the
purposes of calculating the transit probabilities of the extra-solar
planets, we have also assumed that the stellar mass follows the
mass-radius relationship $M_* = R_*^{1.25}$ (\citealt{tingley05}).

It is the 3 quantities $R_*$, $P_{rot}$ and $x$ that constitute the
`proposal parameters' with analogy to the description of the
implementation of MCMC outlined by \cite{cameron07}.  We can then
perform a random walk through parameter space by perturbing each
proposal parameter from its previous value by a random amount:

\[ \,\,\,R_{*,i} = R_{*,i-1} + G \sigma_R \]
\[ P_{rot,i} = P_{rot,i-1} + G \sigma_P \]
\[ \,\,\,\,\, \, \, \,x_{i} = x_{i-1} + G \sigma_x, \]

\noindent where $G$ is a Gaussian random number with zero mean and
unit variance. The initial value of $x = \cos i$ was set to 0.5 and
given an arbitrary standard deviation $\sigma_x$ = 0.05 which was
later re-evaluated empirically from the Markov chains themselves (see
later).

After each perturbation, $\chi^2$ was evaluated for the new set of
proposal parameters via:

\[ \chi _i ^2 = \frac{(R_{*,i}
  - R_{*,0})^2}{\sigma_R^2} + \frac{(P_{rot,i} -
  P_{rot,0})^2}{\sigma_P^2} \]
\begin{equation}
+ \frac{(2 \pi R_{*,i} \sqrt{\left[ 1 - x_i^2 \right]} / P_{rot,i} - v
  \sin i)^2}{\sigma_v^2},
\end{equation}

\noindent where $ \sqrt{ \left[ 1 - x_i^2 \right]} = \sin i$, $v \sin
i$ is the measured projected stellar rotation velocity and $R_{*,0}$,
$P_{rot,0}$ are the measured stellar radius and rotation period,
respectively. For each jump, if $\chi_i^2 < \chi_{i-1}^2$ then the new
parameters were accepted, otherwise the new parameters were accepted
with the acceptance probability given by $\exp \left[ - (\chi_i^2 -
  \chi_{i-1}^2) / 2 \right]$ (the Metropolis--Hastings rule). The
uncertainty $\sigma_x$ was recomputed from the Markov chains
themselves every 100 successful steps by calculating the standard
deviation on $x$ over these 100 jumps.

We found that it was necessary to carry out 1,000,000 jumps in order
for the MCMC to return the maximum likelihood value of $\sin i$ that
accurately approached the value obtained from
equation~\ref{eqn:sini}. The Markov chains were then evaluated
(after discarding a 1000-step long burn-in phase) in
order to determine the 1-$\sigma$ two-tailed confidence limits on
$\sin i$. In addition, for each set of new parameters generated within
the Markov chain, we evaluated whether or not the extra-solar planet
(or extra-solar planets in the case of multiple systems) would transit
the host star. Thus our implementation of MCMC also returns the
transit probability for each extra-solar planet in the study. We
should note, however, that we have assumed that the extra-solar
planets follow circular orbits, so our calculated transit
probabilities may not be accurate for extra-solar planets with highly
eccentric orbits. Furthermore, objects are flagged as transiting if the
planets centre crosses the stellar disc -- the planetary radius is
not taken into account. The results of the MCMC analysis are shown in
Table~\ref{tab:3}.

\section{Possible sources of systematic errors}
\label{sec:syserrs}

While we have already highlighted possible sources of error arising
from, for example, variation of the chromospheric emission due to
rotation of active regions or stellar activity cycles, it is pertinent
to look into other possible sources of systematics. These include
potential biases as a result of differing line-of-sight effects, our
use of an inhomogeneous set of data from a number of different
studies, selection effects, and problems arising due to our ignorance
of the physics at work that affect the measurables in
equation~\ref{eqn:sini}. We shall discuss possible systematics
affecting the estimation of the parameters in the right-hand side of
equation~\ref{eqn:sini} (namely $P_{rot}$, $v \sin i$ and $R_*$) in
turn.

\subsection{Systematic errors on $P_{rot}$}

Most of the stellar rotation periods reported in this paper have been
estimated from the strength of the chromospheric Ca II H \& K emission
with the exception of a few that have been determined
photometrically. Rotation periods calculated from analysis of Ca II H
\& K emission are, as previously described in detail in
section~\ref{sec:rhk_data}, impacted by variability caused by activity
cycles and the temporal evolution of magnetic regions. On top of this,
however, there may also be line-of-sight geometry effects to consider
for given starspot or active region distributions.  For instance,
Doppler images of rapidly rotating active stars (e.g. \citealt{skelly09};
\citealt{watson07}; \cite{watson06}; \citealt{cameron02})
have revealed the presence of high-latitude and even polar spots
covering a significant fraction of the stellar surface. This is in
stark contrast to our Sun where spots are rarely observed at latitudes
$>$40$^{\circ}$ and seldomly cover more that $\sim$ 1 per cent of the
solar surface.

Assuming that the bulk Ca II H \& K emission arises from regions
associated with starspots, then the distribution of spots coupled with
the inclination of the stellar rotation axis to the observers
line-of-sight could systematically affect the derived rotation
period. For example, consider a star with a large polar active
region. In this case the observer would see a larger projected area of
activity when viewed at a low inclination (from above the pole)
compared to the same distribution viewed edge-on at high inclinations.
Under this scenario this would lead to seemingly higher levels of
chromospheric activity observed in rapidly rotating stars viewed at
low inclinations. This, in turn, would lead to systematically shorter
$P_{rot}$ estimates for rapidly-rotating, low inclination stars and
(from equation~\ref{eqn:sini}) drive the estimated $\sin i$ to even
lower values. Conversely, rapidly rotating stars viewed at high
inclinations would, presumably, have $\sin i$ estimates systematically
biased towards higher values. Unfortunately, we are largely ignorant
of the exact interplay between spot numbers, sizes and distributions
and the corresponding Ca II H \& K emission which makes the estimation
of the magnitude of this effect beyond the scope of this paper. This
is further exasperated by our lack of detailed understanding of how
stellar activity varies as a function of spectral-type and stellar age
(or, equivalently, rotation rate).

In addition, the majority of exoplanet host stars are, by selection,
relatively inactive and therefore exhibit low Ca II H \& K
emission. For these stars there may be a possible bias towards
measurements of higher $R'_{hk}$ values, since it should be easier to
detect their Ca II H \& K emission at the peak of their activity. This
would cause the estimated stellar rotation rates to be too fast,
skewing our $\sin i$ distribution to low values.

\subsection{Systematic errors on derived $v \sin i$'s}

The $v \sin i$ values quoted in this work come from a variety of
sources and are not from a homogeneous sample. For many exoplanet
discovery papers the value of the rotational broadening of the host
star is often reported with little discussion as to how this was
determined.  This is of little surprise, since the authors are largely
preoccupied with characterising the planet rather than the parent
star. However, it raises the question of whether the reported $v \sin
i$ values are accurate and, in addition, also correct relative to one
another.

The observed stellar line broadening is a function of the intrinsic
line-profile width, convolved with the rotationally broadened profile
and the instrumental profile. Thus, to first approximation the
observed line-profile full-width at half maximum ($\Delta_{obs}$) is
given by

\begin{equation}
\Delta_{obs} = \sqrt{\left(\alpha \times v\sin i \right)^2 + \xi^2 +
  \Delta_{inst}^2},
\label{eqn:fwhm}
\end{equation}

\noindent where $\alpha$ is an arbitrary scaling constant to convert
$v \sin i$ to a full-width half maximum, $\xi$ in the intrinsic
line-profile full-width half maximum, and $\Delta_{inst}$ is the
instrumental profile.  If the instrumental profile and/or intrinsic
line-profile are ignored then the derived $v \sin i$ will be an
overestimate. This would drive the $\sin i$ distribution towards high
values. Furthermore, this systematic bias would be more profound for
slow rotators and also for systems seen at low inclinations. If the
intrinsic line-profile, $\xi$, is not properly treated in the
estimation of $v \sin i$ then, since hotter stars have broader
intrinsic line-profile widths, the problem will also become
progressively worse for earlier spectral-type stars.  Clearly many of
these potential systematic biases could be alleviated if the data were
taken from a homogeneous set and analysed in a consistent manner.

\subsection{Errors on $R_*$}

Most of the stellar radii presented in this work have been calculated
by comparison of theoretical stellar atmosphere models to observed
high-resolution spectra. As outlined by  (\citealt{brown10}), this
yields small formal errors on the radius (often better than 2 per
cent), but is heavily model dependent.  \cite{brown10} compared the
results of this technique with a group of well calibrated eclipsing
binaries, as well as single stars for which good fundamental
parameters were known from asteroseismology investigations.  While the
results of the models compare accurately with the slowly rotating,
inactive, single stars in the asteroseismic sample, a discrepancy
occurs when applied to the stellar components in the eclipsing binary
sample.  Indeed, for this sample a mass-dependent underestimate of the
stellar radius by $\sim$ 4 per cent for low-mass stars and which
gradually decreased, becoming negligible for stars with masses above
$\sim$1.4 $M_{\odot}$, was found.

The explanation for this underestimation  is that the more rapidly
rotating active stars have their radii inflated due to blocking of
energy transport in the outer convection regions by star spots. Since
spots do not affect the core luminosity, the stars response to the
appearance of spots is to inflate the stellar radius and/or increase
the temperature of the non-spotted regions in the photosphere. Thus,
more rapidly rotating stars in our sample are likely to have their
radii underestimated, leading to a skew to high $\sin i$'s.  Given
that most of exoplanet host stars are (by selection) slowly rotating,
we don't expect this to be a dominant source of systematic error.
There are, however, a few cases where stars have several radii
estimates available in the literature from different sources which
differ quite dramatically. We are unable to offer any reasonable
explanation for these discrepancies (highlighted in
Section~\ref{sec:technique}).

\subsection{Non-Gaussianity of errors}

In the Markov-chain Monte Carlo analysis performed in
Section~\ref{sec:mcmc} we have assumed that the errors on the stellar
radius, rotation period and $v \sin i$ measurements are Gaussian in
nature. This assumption, however, may not be true, especially given
the range of systematic errors that may exist as discussed
above. While we could, technically, inject non-Gaussian errors and
assume modified probability distributions for each of the parameters
in our MCMC analysis, any such probability distribution would have to
be guessed at.  We feel that, given the complexity and interplay
arising due to the systematics discusses above, any such attempt might
be just as misleading as our assumption of Gaussianity.

\section{Transiting systems and transiting probabilities}

The transiting planets included in our literature search are
summarised in Table~\ref{tab:trans} and provide a good test of how
accurate our method is, since all these systems should have $\sin i
\sim 1$. Indeed, 6 out of the 11 transiting systems have $\sin i$'s
$>$ 0.9, and 10 out of the 11 are within 2$-\sigma$ of $\sin i = 1$.
The notable exception is OGLE-TR-111, which yields a wildly discrepant
value of $\sin i$ = 4.518, probably due to systematic errors in measuring the
stellar parameters due to its faintness (see Appendix A for more
details). This probably also explains why we obtain a relatively low
$\sin i$ = 0.763 for OGLE-TR-113. In addition, the extra-solar planet
host star HAT-P-1 gives a low $\sin i$ = 0.747, but in this case it is
actually the member of a binary system and no $B-V$ value is available
for the individual host star. We calculated a $B-V$ value using
$T_{eff}$ = 5975K from \cite{bakos07a} and the relationship
$\log{T_{eff}}$ = 3.908 - 0.234 ($B-V$) from \cite{noyes84}. It is,
therefore, probable that the rotation period we have calculated from
$\log R'_{HK}$ and our estimated ($B-V$) colour is incorrect. Finally,
we find a low $\sin i$ of 0.754$^{+0.177}_{-0.165}$ for the transiting
system HD 17156. This infers a misalignment angle between the
spin-axis of the host star and the orbit of the planet of
41$^{\circ}$$^{+13}_{-21}$. We note that this is consistent with the
misalignment angle of 62$^{\circ} \pm$ 25$^{\circ}$ measured by
\cite{narita08} from the Rossiter-McLaughlin effect. This, however,
has been more recently revised to 9.4$^{\circ} \pm$ 9.3$^{\circ}$ by
\cite{cochran08}. It would be interesting to confirm these
observations.

The remainder of the transiting extra-solar planet host stars,
however, all yield $\sin i$'s close to 1, with the TrES candidates
providing particularly encouraging results. It is comforting to find
that 8 of the known transiting extra-solar planets in our sample
(excluding OGLE-TR-111b, HAT-P-1b and HD 17156 for the reasons
outlined earlier) lie within the top 20 transiting candidates as
determined from our MCMC analysis. Furthermore, the technique flagged
the known transiting extra-solar planet OGLE-TR-56b as the most likely
to transit. This suggests that the use of MCMC could be an efficient
tool in identifying extra-solar planet transit candidates from
spectroscopic analysis of the host stars.

In Table~\ref{tab:prob} we have listed the top 20 spectroscopically
discovered extra-solar planets with the highest transit probabilities
as determined from the MCMC analysis. Naturally there is a bias for
extra-solar planets with short orbital periods to be flagged as more
probable transit candidates on account of their close proximity to the
host star. This means that any long-period extra-solar planet that has
a relatively high transit probability is worthy of mention, since such
planets are more likely to have been overlooked in targeted transit
searches. From Table~\ref{tab:prob}, HD117176b is perhaps the most
interesting candidate. With an orbital period of 116.689 days it would
be of no surprise if transits had been missed.

\section{Results}
\label{sec:results}

For the purposes of this paper, we have adopted the Working Group on
Extra-solar Planets definition of a planet to be an object below the
limiting mass for thermonuclear fusion of deuterium, currently
calculated to be 13$M_J$. It is comforting, therefore, to find that
only 6 extra-solar planet candidates in our sample have calculated
masses that place them over this deuterium burning limit. These are HD
81040b (17.1 $M_J$), HD 136118b (14.5 $M_J$), HD 141937b (17.6 $M_J$),
HD 162020b (147.8 $M_J$), HD 168443c (18.1 $M_J$) and HD 202206b (17.7
$M_J$). Of these 6, HD 168443c and HD 202206b already had minimum
masses calculated to be $>$ 17.4 $M_J$. Of the remainder, only HD
162020b has a revised mass that puts it significantly above the 13
$M_J$ cut-off for planetary status and, with a calculated true mass of
148 $M_J$, we suggest that the companion is most likely an $\sim$M4
dwarf. Including the errors on $\sin i$, we find a possible minimum
mass (at the 1-$\sigma$ level) of 67 $M_J$, and thus the possibility
of a brown dwarf companion cannot be ruled out.  We believe that a
companion mass much larger than 148 $M_J$ is unlikely since it would
have a clear spectral signature.  Interestingly, \cite{udry02} use
tidal dissipation arguments to conclude that the companion to HD
162020 is probably a brown dwarf, although they could also not rule
out a low-mass star, in agreement with our results.

\begin{figure}
\psfig{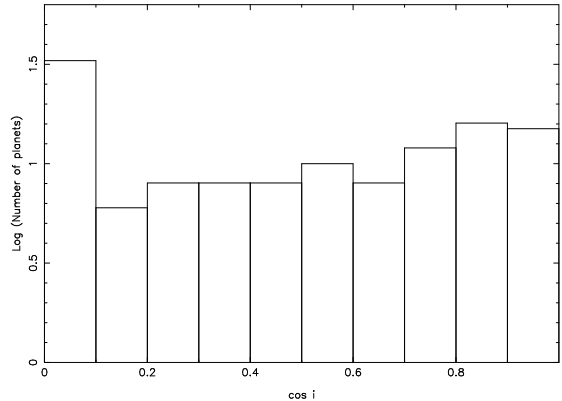}
\caption{A logarithmic histogram of the $\cos i$ values for the
spectroscopically-discovered extra-solar planet systems in our sample.}
\label{fig:sinihist}
\end{figure}

Fig.~\ref{fig:sinihist} shows a histogram of the $\cos i$ values
obtained from the MCMC analysis for the spectroscopically-discovered
systems in our catalogue. This shows a peak at high inclinations
where the systems with naive estimators of $\sin i >$ 1 pile up
at $\sin i$ = 1 in the subsequent MCMC analysis. Given an isotropic
distribution of stellar rotation inclination angles, one would
expect the $\cos i$ distribution to be flat. However, since
the amplitude of a planets' radial velocity signal decreases with
$\sin i$ then we would expect planet detectability to also drop
off towards low $\sin i$. There does, however, seem
to be a slight excess of low inclination systems, with a general decrease
in the number of systems populating higher inclinations (ignoring the
pile-up). We interpret this overall shape of the distribution to be
due to systematic errors pushing high and moderately inclined stars
into the $\cos i = 0$ `spike'. Indeed, one could envisage
redistributing the $\cos i \sim 0$ systems to lower inclinations,
thereby flattening out the observed distribution.

This gives us some confidence that our rejection of stars with naive
$\sin i$ estimates greater than 1-$\sigma$ above $\sin i = 1$ is reasonable.
Inclusion of more objects with naive estimates of $\sin i > 1$ in the
MCMC analysis would simply produce a large number of systems with $\sin i$
very close to 1 and very small $\sin i$ uncertainties on account
of enforcing our prior knowledge that $\sin < 1$. For these reasons,
inclusion of these objects would be questionable as it is
likely that the errors have been underestimated in for these objects, or
they are affected by systematics.

\begin{figure*}
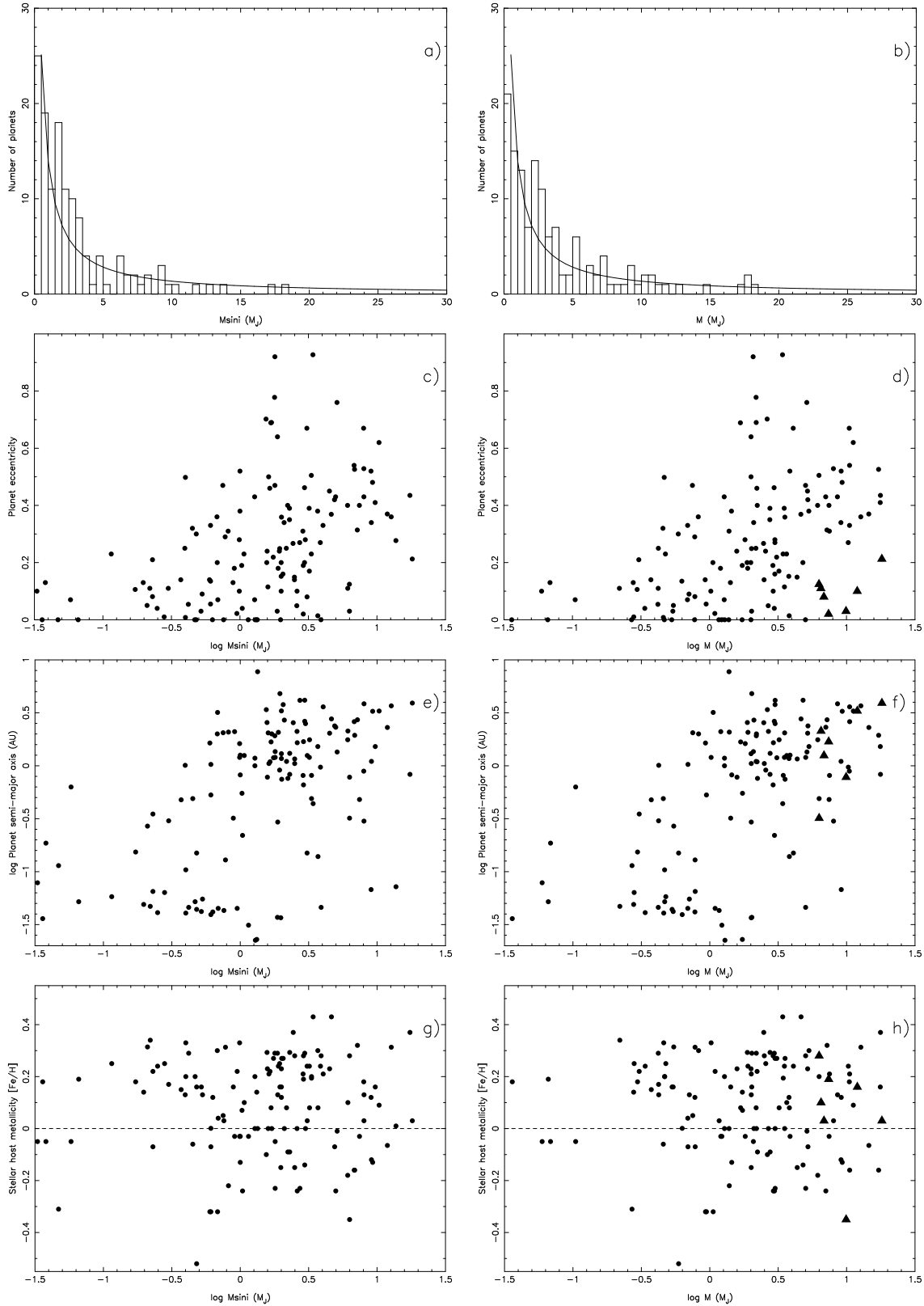

\begin{tabular}{ll}
\psfig{figure=planet1.ps,width=7.3cm,angle=-90} &
\psfig{figure=planet2.ps,width=7.3cm,angle=-90} \\
\psfig{figure=planet3.ps,width=7.3cm,angle=-90} &
\psfig{figure=planet4.ps,width=7.3cm,angle=-90} \\
\psfig{figure=planet5.ps,width=7.3cm,angle=-90} &
\psfig{figure=planet6.ps,width=7.3cm,angle=-90} \\
\psfig{figure=planet7.ps,width=7.3cm,angle=-90} &
\psfig{figure=planet8.ps,width=7.3cm,angle=-90} \\
\end{tabular}
\caption{Top panel: a histogram of the number of extra-solar planets
with a) observed minimum masses $M \sin i$ and (b) their calculated true
masses, $M$ (both in units of Jupiter masses, M$_J$). The solid line
indicates a mass distribution characterised by the power-law $dN/dM \propto
M^{-1.1}$. Figures c) and d) plot the orbital eccentricity versus
the minimum extra-solar planet mass and their calculated true masses,
respectively. A number of relatively high mass, low eccentricity ($e <
0.25$) planets discussed in Section~\protect\ref{sec:results} have
been indicated using triangular
markers. Figures~\protect\ref{fig:plots}e and f are the same as c) and
d), but against extra-solar planet semi-major axis. Finally, figures g)
and h) plot the host star metallicity [Fe/H] versus $M \sin i$ and true mass.
The horizontal dashed line represents solar metallicity.}
\label{fig:plots}
\end{figure*}

A summary of our findings are presented in Fig.~\ref{fig:plots}, which
shows both the minimum extra-solar planet masses and `true' masses
versus properties such as number frequency, orbital semi-major axis,
orbital eccentricity and host star metallicity. In order to make the
comparison fair, we only plot the minimum extra-solar planet masses
for those planets which have been included in the MCMC analysis
(i.e. only those systems presented in Table~\ref{tab:3}).

Comparing the results of the minimum and true extra-solar planet
masses versus number frequency (top panel, Fig.~\ref{fig:plots}), we
still find that lower mass extra-solar planets are more common, with a
tail of high-mass companions. This mass distribution can be roughly
characterised by the power-law $dN/dM \propto M^{-1.1}$
(\citealt{butler06}), and does not change appreciably once the $\sin
i$ dependency has been removed. This has previously been noted in a
purely statistical analysis of extra-solar planet masses by
\cite{jorissen01}. It is often cited (e.g. \citealt{jorissen01}) that
the number of planets with minimum masses above 10 $M_J$ is
essentially zero -- suggesting that planetary formation is a distinct
process from that which forms low-mass and sub-stellar (e.g. brown
dwarf) objects. When considering their true masses, the planet
frequency appears to drop to zero around a slightly higher limit of
$\sim$13 $M_J$. Interestingly, this corresponds to the adopted upper
mass-limit for a planet at the planet/brown dwarf boundary. Given the
low number of extra-solar planets in this mass range, however, it is
difficult to definitively place a higher mass `cut-off' for
extra-solar planets.

Figs.~\ref{fig:plots}c \& d show the extra-solar planet minimum
masses and true masses versus orbital eccentricity, respectively. It
can be seen that when considering just the extra-solar planet minimum
masses, there is a dearth of low eccentricity ($e < 0.2$) extra-solar
planets for minimum masses greater than $\sim$ 6$M_J$, as already
noted by several other authors (e.g. \citealt{butler06}). When one
considers the true masses, however, we find 6 extra-solar planets with
masses in the range $6 - 12M_J$, along with one brown dwarf companion
(all indicated by triangular markers), with $e \sim 0.2$.

These high-mass, low-eccentricity (hereafter HMLE) planets are also
indicated in Fig.~\ref{fig:plots}f which plots semi-major axis
versus true extra-solar planet mass. We find that these HMLE extra-solar
planets have a wide range of semi-major axes, including one that has
one of the largest semi-major axes included in our sample. Therefore,
one presumably cannot just appeal to orbital circularisation through
tidal forces due to the planet's close proximity to the host star to
explain these HMLE extra-solar planets.

We have applied the Hartigan dip-test (\citealt{hartigan85}) to check
for the non-unimodality of the distribution of orbital eccentricities
for exoplanets with masses greater than 5$M_J$, rejecting objects
above 13$M_J$. This returns a 55 per
cent probability that the eccentricity distribution is not unimodal
and may, therefore, be indicative of two different populations of
exoplanets. A larger sample of extra-solar
planets in the $>5 M_J$ mass range is needed before any firm conclusions
about the significance of these HMLE extra-solar planets can be drawn. A larger
sample of high mass extra-solar planets would also help to establish
whether the gap in orbital eccentricities between $e$ = 0.2 -- 0.3 for
high mass extra-solar planets apparent in Fig.~\ref{fig:plots}d is
real. If confirmed, however, the presence of these HMLE extra-solar
planets, and the gap in orbital eccentricities between $e$ = 0.2 --
0.3, hints at a distinct evolution and/or formation process for these
extra-solar planets.

Studies of brown dwarfs and spectroscopic binaries have shown that
they exhibit a similar eccentricity distribution to the higher-mass
extra-solar planets (extra-solar planets exhibit a trend of increasing
mean orbital eccentricity with increasing mass, as mentioned
earlier). This has led \cite{ribas07} to suggest that the
eccentricity-mass distribution of extra-solar planets may provide a
signature of different extra-solar planet formation mechanisms. They
hypothesize that there are two formation scenarios for extra-solar
planets. The first is that the low-mass population forms by gas
accretion onto an ice-rock core within the circumstellar disk, and
initially form in circular orbits and grow their eccentricities by
varying amounts later. The second is that the high-mass population
forms directly from fragmentation of the pre-stellar cloud (in the
same manner as brown dwarfs and binaries) and would initially be
located in far larger orbits. The subsequent long-distance migration
required to bring them to their current positions is then postulated
to drive these higher mass extra-solar planets to much larger
eccentricities.

If \cite{ribas07} are correct then this might suggest that the
candidates we have identified as HMLE extra-solar planets in
Fig.~\ref{fig:plots} have formed along the same route as the low-mass
planets, i.e.  through gas accretion onto a rock-ice core rather than
via fragmentation. In order to form such massive planets by gas
accretion, we might expect the host stars to have higher
metallicities. Figs.~\ref{fig:plots}g \& h show host star metallicity
$[$Fe/H$]$ versus $M \sin i$ and true mass, respectively, with the
HMLE extra-solar planets indicated by triangles. We note that 5 of the
HMLEs are indeed around host stars with high metallicities but the
remaining HMLE candidate happens to be around one of the most metal
poor host stars in our selection. The anonymous referee has pointed
out that the conclusion that the HMLEs should have higher
metallicities is not the only possibility, and that formation in a
high-mass disc could supply the right environmental conditions as well.
Obviously, the true masses of more extra-solar planets need to be calculated
before any sound conclusions as to whether these HMLEs truly
constitute a distinct population, and the clues they may give us about
planetary formation, can be made.

\section{discussion and conclusions}
\label{sec:disc}

Under the assumption that the rotation axes of extra-solar planet host
stars are aligned perpendicularly to the planes of the extra-solar
planetary orbits, we have used measurements of $R_*$, $v \sin i$ and
$P_{rot}$ to remove the $\sin i$ dependency from 133
spectroscopically-determined extra-solar planet mass
determinations. We find that, bar two problematic cases, the
inclination angles of all the known transiting extra-solar planets in
our sample are commensurate with $\sin i$ = 1, as expected. Using a
Markov-chain Monte Carlo analysis, we have also computed the transit
probabilities of all 133 extra-solar planets from purely spectroscopic
measurements.  We find that all 8 known transiting extra-solar planets
with reliable parameter determinations lie in the top 20 most probable
transiting candidates.  This gives us some confidence that not only
can the technique outlined in this paper be used to correctly estimate
the true masses of extra-solar planets, but also that MCMC can
reliably identify extra-solar planet transit candidates from
spectroscopic measurements.

We find that only 6 out of the 133 extra-solar planets have masses
that place them over the standard 13$M_J$ upper limit for planets,
which indicates that the vast majority of extra-solar planet
candidates found by spectroscopic means are truly planetary in
nature. We also find evidence for a population of high-mass
extra-solar planets with low orbital eccentricities that is not
apparent when only extra-solar planet minimum masses are
considered. It is possible that these extra-solar planets may have
formed along a different path to the other high-mass extra-solar planets.
This suggests that, while some high-mass planets may well form through
fragmentation resulting in high eccentricity orbits as suggested by
\cite{ribas07}, not all high mass planets form in this way.

Only by calculating the true masses of more extra-solar planets can
such distributions, and their impact on our understanding of both
planet {\em and} brown dwarf formation, be properly studied. With 453
extra-solar planet candidates, there are still over 300 extra-solar
planets for which we could not find the necessary data to determine
$\sin i$, or for which the data were unreliable and yielded $\sin i$'s
significantly greater than 1. There are several observational problems
to overcome. In order to calculate the rotation period of the star we
generally must rely on measurements of the strength of the
chromospheric Ca {\sc ii} H \& K lines and apply the
chromospheric-emission / rotation law of \cite{noyes84}.  The
\cite{noyes84} relation has obvious drawbacks (i.e. it is not a direct
measurement of the stellar rotation period), and the Ca {\sc ii} H \& K
emission in these stars may be variable over long-time scales due to,
for example, magnetic activity cycles like the 11-year solar
cycle. Thus measurements of Ca {\sc ii} H \& K need to be averaged
over a suitably long time-span in order to derive a reliable
rotation-period. Whilst we are in the fortunate position that large Ca
{\sc ii} H \& K surveys like the Mt. Wilson survey have observed many
extra-solar planet host stars for several decades now, there are still
many host stars where only one brief `snapshot' of the chromospheric
emission is available from the planet discovery paper. We plan to
commence the targeted monitoring of chromospheric emission from
extra-solar planet host stars, not only to obtain a long-term average
of the chromospheric emission from these stars, but also to see if
variations in the indicators over the {\em actual} rotation period of
the star can be identified. This would give a direct measure of the
stellar rotation period.

The next observational problem is the determination of the projected
stellar equatorial velocity, $v \sin i$. Again, the nature of
the hunt for extra-solar planets means that the host stars are almost
always observed with high-resolution echelle spectrographs from which
the line-broadening can be measured. Due to the low (typically $\sim$
2 km s$^{-1}$) rotation velocities of these stars, rotational
broadening is no longer the dominant line-broadening mechanism, and
other mechanisms such as thermal broadening and turbulence need to be
taken into account.  Many of the quoted $v \sin i$ measurements in the
literature do not fully account for these effects, which require the
use of stellar atmosphere models to estimate the true level of
broadening due to rotation. We plan to systematically analyse the
spectra of extra-solar planet host stars to produce accurate $v \sin
i$ measurements taking into account other broadening mechanisms.

Finally, we note that the inclination of the rotation axis of stars
can be measured using asteroseismology.  \cite{gizon03}
present a technique which determines the stellar axial inclination
from observations of low-degree non-radial oscillations which are
strong functions of $i$. They find that the inclination angle can be
measured using this method to within $\sim 10^{\circ}$ when
$i > 30^{\circ}$. One condition for this technique to work, however,
is that the star must have a high rotation rate, and this restricts
the technique to stars that rotate at least twice as fast as the
Sun. Since the host stars of extra-solar planets are generally slow
rotators (selected in order to avoid `jitter' in the radial velocity
measurements caused by magnetic activity which is enhanced for more
rapidly rotating stars), this technique will not be able to access a
substantial portion of these stars. We therefore believe that, for the
foreseeable future at least, the technique outlined in this paper will
remain the main way in which to remove the $\sin i$ degeneracy in
spectroscopically-determined extra-solar planet masses.

\onecolumn

\begin{center}

\begin{longtable}{lllr@{.}lr@{.}lr@{.}lr@{.}lr@{.}lr@{.}lr@{.}lr@{.}lr@{.}l}

\caption[]{Published data on the properties of 154 extra-solar planet
host stars.  Columns 1 and 2 give the HD and HIP catalogue number of
the host star, respectively, and column 3 gives any other common name
that the star may be known as. Published $v \sin i$ measurements and
the associated error bar, $\sigma_v$, are given in columns 4 and 5,
respectively. Column 6 lists the measured $\log R'_{HK}$ found from
the literature, and columns 7 and 8 list any stellar rotation periods
and corresponding errors that are quoted. Note that the stellar
rotation period may not correspond to the $\log R'_{HK}$ on the same
line. Actual {\em observed rotation periods} are indicated with an
asterisk next to the measurement. The final two columns give the
published values and error bars for the stellar radius.
References for the values are indicated by the
numbers in superscript and can be found at the end of the table.  We
have also included the radii we have calculated for each star from the
Barnes-Evans relationship (reference number 93). Where rotation
periods do not have an associated reference number, they have been
calculated using the adjacent $P_{rot}$ value and the \cite{noyes84}
chromospheric emission -- rotation period relationship along with
$(B-V)$ values taken from the NStED database.}

\\

\hline  & & \multicolumn{1}{c}{Alternative} & \multicolumn{2}{c}{$v
\sin i$} & \multicolumn{2}{c}{$\sigma_v$} & \multicolumn{2}{c}{$\log
R'_{HK}$} & \multicolumn{2}{c}{$P_{rot}$} &
\multicolumn{2}{c}{$\sigma_P$} & \multicolumn{2}{c}{$R_*$} &
\multicolumn{2}{c}{$\sigma_R$} \\ \multicolumn{1}{c}{HD} &
\multicolumn{1}{c}{HIP} & \multicolumn{1}{c}{Name} &
\multicolumn{2}{c}{(km s$^{-1}$)} & \multicolumn{2}{c}{} &
\multicolumn{2}{c}{} & \multicolumn{2}{c}{(days)} &
\multicolumn{2}{c}{} & \multicolumn{2}{c}{($R_{\odot}$)} \\
\multicolumn{1}{c}{(1)} & \multicolumn{1}{c}{(2)} &
\multicolumn{1}{c}{(3)} & \multicolumn{2}{c}{(4)} &
\multicolumn{2}{c}{(5)} & \multicolumn{2}{c}{(6)} &
\multicolumn{2}{c}{(7)} & \multicolumn{2}{c}{(8)} &
\multicolumn{2}{c}{(9)} & \multicolumn{2}{c}{(10)} \\ \hline
\endfirsthead

\multicolumn{21}{c} {{\bfseries \tablename\ \thetable{} -- continued}}
\\ \hline

& & \multicolumn{1}{c}{Alternative} & \multicolumn{2}{c}{$v \sin i$} &
\multicolumn{2}{c}{$\sigma_v$} & \multicolumn{2}{c}{$\log R'_{HK}$} &
\multicolumn{2}{c}{$P_{rot}$} & \multicolumn{2}{c}{$\sigma_P$} &
\multicolumn{2}{c}{$R_*$} & \multicolumn{2}{c}{$\sigma_R$} \\

\multicolumn{1}{c}{HD} & \multicolumn{1}{c}{HIP} &
\multicolumn{1}{c}{Name} & \multicolumn{2}{c}{(km s$^{-1}$)} &
\multicolumn{2}{c}{} & \multicolumn{2}{c}{} &
\multicolumn{2}{c}{(days)} & \multicolumn{2}{c}{} &
\multicolumn{2}{c}{($R_{\odot}$)} \\

\multicolumn{1}{c}{(1)} & \multicolumn{1}{c}{(2)} &
\multicolumn{1}{c}{(3)} & \multicolumn{2}{c}{(4)} &
\multicolumn{2}{c}{(5)} & \multicolumn{2}{c}{(6)} &
\multicolumn{2}{c}{(7)} & \multicolumn{2}{c}{(8)} &
\multicolumn{2}{c}{(9)} & \multicolumn{2}{c}{(10)} \\ \hline \endhead

142........&522&&10&350$^3$&0&500&-5&020$^1$&10&747&.&.&1&440$^4$&0&070 \\
&&&.&.&.&.&-4&950$^{92}$&10&058&.&.&1&353$^{16}$&0&024 \\
&&&.&.&.&.&.&.&.&.&.&.&1&452$^{93}$&0&034 \\
1237......&1292&GJ~3021&6&000$^9$&.&.&-4&440$^5$&10&400$^5$&.&.&0&940$^4$&0&050 \\
&&&5&500$^8$&.&.&-4&270$^6$&12&600$^7$&.&.&0&850$^3$&.&. \\
&&&5&030$^3$&.&.&-4&340$^1$&6&549&.&.&0&865$^{65}$&0&055 \\
&&&.&.&.&.&.&.&.&.&.&.&0&902$^{93}$&0&016 \\
2039......&1931&&3&250$^3$&0&500&-4&980$^1$&26&247&.&.&1&370$^4$&0&200 \\
\end{longtable}

\end{center}
References: $^1$\cite{saffe05}, $^2$\cite{butler06},
$^3$\cite{fischer05}, $^4$NStED, $^5$\cite{barnes01}, $^6$Coralie,
$^7$Geneva, $^8$Coravel, $^9$\cite{nordstroem04},
$^{10}$\cite{moutou05}, $^{11}$\cite{fracassini01},
$^{12}$\cite{pizzolato03}, $^{13}$\cite{wright04},
$^{14}$\cite{barnes07}, $^{15}$California \& Carnegie Planet Search
Team, $^{16}$\cite{valenti05}, $^{17}$\cite{udry06},
$^{18}$\cite{mayor04}, $^{19}$\cite{perrier03},
$^{20}$\cite{fuhrmann98}, $^{21}$\cite{bernkopf01},
$^{22}$\cite{fischer07}, $^{23}$\cite{fischer01},
$^{24}$\cite{saar97}, $^{25}$\cite{reiners03}, $^{26}$\cite{jones06},
$^{27}$\cite{otoole07}, $^{28}$\cite{johnson06a},
$^{29}$\cite{galland05}, $^{30}$\cite{acke04}, $^{31}$\cite{santos02},
$^{32}$\cite{fischer02}, $^{33}$\cite{hatzes06},
$^{34}$\cite{demedeiros99}, $^{35}$\cite{lovis06},
$^{36}$\cite{lowrance05}, $^{37}$\cite{messina01},
$^{38}$\cite{henry96}, $^{39}$\cite{udry03}, $^{40}$\cite{naef04},
$^{41}$\cite{naef01}, $^{42}$\cite{sozzetti06},
$^{43}$\cite{bernacca70}, $^{44}$\cite{korzennik00},
$^{45}$\cite{lovis05}, $^{46}$\cite{fuhrmann97}, $^{47}$\cite{naef07},
$^{48}$\cite{ge06}, $^{49}$\cite{melo07}, $^{50}$\cite{sato03},
$^{51}$\cite{fischer06}, $^{52}$\cite{vogt02}, $^{53}$\cite{udry02},
$^{54}$\cite{eggenberger06}, $^{55}$\cite{soderblom82},
$^{56}$\cite{benz84}, $^{57}$\cite{bakos07b}, $^{58}$\cite{dasilva06},
$^{59}$\cite{santos04}, $^{60}$\cite{pepe02}, $^{61}$\cite{johnson07},
$^{62}$\cite{johnson06b}, $^{63}$\cite{bouchy05},
$^{64}$\cite{melo06}, $^{65}$\cite{masana06}, $^{66}$\cite{naef03},
$^{67}$\cite{henry02}, $^{68}$\cite{santos00}, $^{69}$\cite{mazeh00}
$^{70}$\cite{fuhrmann98b}, $^{71}$\cite{locurto06},
$^{72}$\cite{pepe04}, $^{73}$\cite{alonso04},
$^{74}$\cite{sozzetti04}, $^{75}$\cite{narita07},
$^{76}$\cite{laughlin05}, $^{77}$\cite{sozzetti07},
$^{78}$\cite{bakos07a}, $^{79}$\cite{pont07}, $^{80}$\cite{santos06},
$^{81}$\cite{queloz00}, $^{82}$\cite{bouchy04},
$^{83}$\cite{konacki04}, $^{84}$\cite{torres08},
$^{85}$\cite{konacki05}, $^{86}$\cite{irwin08}, $^{87}$\cite{henry02b},
$^{88}$\cite{strassmeier00b}, $^{89}$\cite{butler03},
$^{90}$\cite{butler00}, $^{91}$\cite{santos01},
$^{92}$\cite{jenkins06},
$^{93}$ Derived from the Barnes-Evans relationship of \cite{fouque97}

\twocolumn

\onecolumn

\begin{center}

\begin{longtable}{lr@{.}lr@{.}lr@{.}lr@{.}lr@{.}lr@{.}lr@{.}lr@{.}lr@{.}lcc}
\caption[]{Adopted parameters and $\sin i$ estimates for all extra-solar
planet host stars with $\sin i > 1$ as derived from
equation~\ref{eqn:sini}. The first 3 columns are as
described in Table~\ref{tab:1}. Subgiants are indicated with an asterisk
as they may not follow the rotation period -- activity relationship
of \cite{noyes84}. Column 4 lists the stellar rotation
period (in days) obtained from the measured $\log R'_{HK}$'s listed in
Table~\ref{tab:1}, and column 5 gives the error bar adopted from the
scatter measured for the Ca {\sc ii} H \& K emission -- rotation
period relationship of \cite{noyes84}. Columns 6 and 7 give the radii
and associated error bar adopted from Table~\ref{tab:1}. See
Section~\ref{sec:appl} for an in-depth discussion of how the adopted
values were obtained. The final two columns give the resulting $\sin
i$ value and corresponding error bar which have been calculated using
equation~\ref{eqn:sini} and a formal error propagation. Sub-giants are
indicated with an asterisk.}  \\

\hline 
\multicolumn{1}{c}{HD or} & \multicolumn{2}{c}{$v
\sin i$} & \multicolumn{2}{c}{$\sigma_v$} &
\multicolumn{2}{c}{$P_{rot}$} & \multicolumn{2}{c}{$\sigma_P$} &
\multicolumn{2}{c}{$R_*$} & \multicolumn{2}{c}{$\sigma_R$} &
\multicolumn{2}{c}{$\sin i$} & \multicolumn{2}{c}{$\pm$} \\
\multicolumn{1}{c}{Alt. name} & \multicolumn{2}{c}{(km s$^{-1}$)} &
\multicolumn{2}{c}{} & \multicolumn{2}{c}{(days)} &
\multicolumn{2}{c}{} & \multicolumn{2}{c}{($R_{\odot}$)} \\
\multicolumn{1}{c}{(1)} & \multicolumn{2}{c}{(2)} &
\multicolumn{2}{c}{(3)} & \multicolumn{2}{c}{(4)} &
\multicolumn{2}{c}{(5)} & \multicolumn{2}{c}{(6)} &
\multicolumn{2}{c}{(7)} & \multicolumn{2}{c}{(8)} &
\multicolumn{2}{c}{(9)}  \\
\hline \endfirsthead

\multicolumn{21}{c} {{\bfseries \tablename\ \thetable{} -- continued}}
\\ \hline

\multicolumn{1}{c}{HD or} & \multicolumn{2}{c}{$v \sin i$} &
\multicolumn{2}{c}{$\sigma_v$}& \multicolumn{2}{c}{$P_{rot}$} &
\multicolumn{2}{c}{$\sigma_P$} & \multicolumn{2}{c}{$R_*$} &
\multicolumn{2}{c}{$\sigma_R$} & \multicolumn{2}{c}{$\sin i$} &
\multicolumn{2}{c}{$\pm$} \\

\multicolumn{1}{c}{Alt. name} & \multicolumn{2}{c}{(km s$^{-1}$)} &
\multicolumn{2}{c}{} & \multicolumn{2}{c}{(days)} &
\multicolumn{2}{c}{} & \multicolumn{2}{c}{($R_{\odot}$)} \\

\multicolumn{1}{c}{(1)} & \multicolumn{2}{c}{(2)} &
\multicolumn{2}{c}{(3)} & \multicolumn{2}{c}{(4)} &
\multicolumn{2}{c}{(5)} & \multicolumn{2}{c}{(6)} &
\multicolumn{2}{c}{(7)} & \multicolumn{2}{c}{(8)} &
\multicolumn{2}{c}{(9)} \\ \hline \endhead

142*        & 10 & 349 & 0 & 500 &  10 & 524 &  0 & 599 &  1 & 389 &  0 & 018 & 1 & 548 & 0 & 117 \\
2039*       &  3 & 250 & 0 & 500 &  25 & 487 &  2 &  34 &  1 & 256 &  0 & 097 & 1 & 302 & 0 & 247 \\
3651        &  1 & 149 & 0 & 500 &  44 & 000 &  9 & 793 &  0 & 878 &  0 & 008 & 1 & 137 & 0 & 555 \\
8574        &  4 & 327 & 0 & 386 &  17 & 073 &  0 & 884 &  1 & 383 &  0 & 029 & 1 & 054 & 0 & 111 \\
9826        &  9 & 482 & 0 & 362 &  11 & 910 &  1 &  18 &  1 & 586 &  0 & 017 & 1 & 406 & 0 & 132 \\
11506       &  5 & 000 & 0 & 447 &  18 & 300 &  0 & 696 &  1 & 397 &  0 & 046 & 1 & 293 & 0 & 132 \\
11964*      &  2 & 700 & 0 & 500 &  50 & 492 &  2 & 553 &  2 & 032 &  0 & 045 & 1 & 325 & 0 & 256 \\
13445       &  2 & 259 & 0 & 179 &  27 & 240 &  6 & 203 &  0 & 823 &  0 & 003 & 1 & 477 & 0 & 356 \\
19994       &  8 & 511 & 0 & 408 &  10 & 783 &  1 & 682 &  1 & 698 &  0 & 028 & 1 & 067 & 0 & 175 \\
23127       &  3 & 299 & 0 & 500 &  32 & 034 &  2 & 285 &  1 & 574 &  0 & 071 & 1 & 326 & 0 & 230 \\
23596       &  3 & 956 & 0 & 381 &  21 & 251 &  1 & 108 &  1 & 583 &  0 & 048 & 1 & 049 & 0 & 119 \\
27442*      &   2 & 873 & 0 & 257 &  89 & 184 & 15 & 674 &  4 & 335 &  0 & 427 & 1 & 167 & 0 & 257 \\
27894       &   1 & 500 & 1 & 000 &  44 & 449 &  4 & 177 &  0 & 844 &  0 & 031 & 1 & 559 & 1 & 051 \\
28185       &   2 & 484 & 0 & 461 &  29 & 976 &  2 & 685 &  1 & 062 &  0 & 031 & 1 & 384 & 0 & 288 \\
30177       &   2 & 959 & 0 & 500 &  45 & 399 &  2 & 896 &  1 & 152 &  0 & 028 & 2 & 303 & 0 & 419 \\
33283       &   3 & 360 & 0 & 447 &  58 & 678 &  6 & 985 &  1 & 306 &  0 & 049 & 2 & 981 & 0 & 544 \\
33564       &  12 & 390 & 0 & 937 &   6 & 802 &  0 & 429 &  1 & 503 &  0 & 024 & 1 & 107 & 0 & 110 \\
33636       &   3 & 080 & 0 & 500 &  16 & 697 &  0 & 966 &  1 & 003 &  0 & 022 & 1 & 012 & 0 & 176 \\
38529       &   3 & 899 & 0 & 500 &  37 & 761 &  2 & 210 &  2 & 750 &  0 & 079 & 1 & 057 & 0 & 152 \\
50499       &   4 & 209 & 0 & 500 &  22 & 160 &  1 & 146 &  1 & 428 &  0 & 027 & 1 & 289 & 0 & 168 \\
52265       &   4 & 775 & 0 & 447 &  15 & 791 &  1 & 191 &  1 & 275 &  0 & 022 & 1 & 168 & 0 & 142 \\
63454       &   1 & 899 & 1 & 000 &  20 & 251 &  5 & 316 &  0 & 744 &  0 & 024 & 1 & 021 & 0 & 601 \\
68988       &   2 & 839 & 0 & 500 &  26 & 459 &  0 & 926 &  1 & 182 &  0 & 037 & 1 & 255 & 0 & 228 \\
73526       &   2 & 620 & 0 & 500 &  35 & 643 &  2 & 433 &  1 & 505 &  0 & 077 & 1 & 225 & 0 & 256 \\
75289       &   4 & 139 & 0 & 500 &  16 & 839 &  1 & 201 &  1 & 271 &  0 & 016 & 1 & 083 & 0 & 152 \\
75732       &   2 & 467 & 0 & 447 &  46 & 791 &  3 & 800 &  0 & 953 &  0 & 009 & 2 & 392 & 0 & 475 \\
80606       &   1 & 431 & 0 & 384 &  42 & 254 &  2 &  62 &  0 & 941 &  0 & 192 & 1 & 268 & 0 & 432 \\
86081       &   4 & 200 & 0 & 500 &  24 & 838 &  1 & 683 &  1 & 295 &  0 & 079 & 1 & 590 & 0 & 238 \\
88133*      &   2 & 185 & 0 & 353 &  49 & 838 &  3 & 263 &  2 & 080 &  0 & 113 & 1 & 033 & 0 & 189 \\
99109*      &   1 & 891 & 0 & 447 &  48 & 485 &  3 & 252 &  1 & 081 &  0 & 048 & 1 & 675 & 0 & 418 \\
99492       &   1 & 379 & 0 & 353 &  46 & 585 &  3 & 923 &  0 & 789 &  0 & 031 & 1 & 609 & 0 & 438 \\
100777      &   1 & 800 & 1 & 000 &  40 & 084 &  1 & 433 &  1 & 133 &  0 & 061 & 1 & 258 & 0 & 703 \\
102195      &   3 & 226 & 0 & 069 &  18 & 429 & 10 & 979 &  0 & 835 &  0 & 013 & 1 & 405 & 0 & 838 \\
108874      &   2 & 220 & 0 & 500 &  40 & 610 &  1 & 401 &  1 & 246 &  0 & 070 & 1 & 429 & 0 & 335 \\
109749      &   2 & 399 & 0 & 447 &  27 & 091 &  1 & 810 &  1 & 243 &  0 & 075 & 1 & 032 & 0 & 213 \\
111232      &   2 & 600 & 1 & 000 &  30 & 437 &  2 & 263 &  0 & 899 &  0 & 017 & 1 & 737 & 0 & 681 \\
117176      &   2 & 827 & 0 & 249 &  35 & 463 &  3 &   4 &  1 & 825 &  0 & 025 & 1 & 085 & 0 & 133 \\
128311      &   3 & 649 & 0 & 500 &  10 & 778 &  2 & 714 &  0 & 769 &  0 & 011 & 1 & 009 & 0 & 289 \\
130322      &   1 & 667 & 0 & 447 &  29 & 377 & 19 & 924 &  0 & 824 &  0 & 026 & 1 & 173 & 0 & 856 \\
134987      &   2 & 169 & 0 & 500 &  33 & 778 &  1 & 649 &  1 & 225 &  0 & 018 & 1 & 181 & 0 & 278 \\
142022A     &   2 & 100 & 1 & 000 &  42 & 052 &  2 & 368 &  1 & 085 &  0 & 028 & 1 & 607 & 0 & 772 \\
145675      &   1 & 560 & 0 & 500 &  48 & 500 &  1 & 137 &  0 & 984 &  0 & 009 & 1 & 519 & 0 & 488 \\
149143*     &   3 & 979 & 0 & 447 &  26 & 703 &  2 &  31 &  1 & 487 &  0 & 055 & 1 & 411 & 0 & 198 \\
160691      &   3 & 662 & 0 & 182 &  32 & 157 &  2 & 172 &  1 & 322 &  0 & 018 & 1 & 758 & 0 & 149 \\
164922      &   1 & 808 & 0 & 447 &  44 & 192 &  1 & 547 &  0 & 980 &  0 & 013 & 1 & 610 & 0 & 402 \\
168443      &   2 & 100 & 0 & 447 &  38 & 606 &  0 & 675 &  1 & 595 &  0 & 030 & 1 & 004 & 0 & 215 \\
177830*     &   2 & 540 & 0 & 500 &  65 & 711 &  6 & 921 &  3 & 129 &  0 & 105 & 1 & 053 & 0 & 237 \\
178911B     &   1 & 939 & 0 & 500 &  36 & 250 &  2 &  24 &  1 & 130 &  0 & 183 & 1 & 229 & 0 & 381 \\
185269*     &   5 & 679 & 0 & 447 &  21 & 458 &  1 & 382 &  1 & 890 &  0 & 054 & 1 & 273 & 0 & 134 \\
186427      &   2 & 253 & 0 & 315 &  29 & 343 &  0 & 767 &  1 & 167 &  0 & 009 & 1 & 119 & 0 & 159 \\
187085      &   5 & 099 & 0 & 500 &  14 & 349 &  1 & 206 &  1 & 331 &  0 & 045 & 1 & 085 & 0 & 145 \\
190360      &   2 & 320 & 0 & 447 &  35 & 807 &  0 & 621 &  1 & 151 &  0 & 013 & 1 & 425 & 0 & 276 \\
190647      &   1 & 969 & 0 & 832 &  40 & 977 &  1 & 410 &  1 & 496 &  0 & 061 & 1 & 065 & 0 & 453 \\
192263      &   2 & 501 & 0 & 447 &  20 & 773 & 12 & 259 &  0 & 775 &  0 & 014 & 1 & 324 & 0 & 817 \\
196050      &   3 & 235 & 0 & 447 &  23 & 282 &  7 & 293 &  1 & 321 &  0 & 039 & 1 & 126 & 0 & 387 \\
196885*     &   7 & 750 & 0 & 500 &  12 & 306 &  0 & 672 &  1 & 387 &  0 & 027 & 1 & 358 & 0 & 117 \\
209458      &   4 & 280 & 0 & 367 &  14 & 914 &  0 & 629 &  1 & 150 &  0 & 029 & 1 & 096 & 0 & 108 \\
210277      &   1 & 839 & 0 & 447 &  40 & 141 &  1 & 849 &  1 & 081 &  0 & 012 & 1 & 349 & 0 & 334 \\
212301      &   6 & 220 & 1 & 000 &  11 & 340 &  0 & 492 &  1 & 172 &  0 & 030 & 1 & 188 & 0 & 200 \\
216435*     &   5 & 780 & 0 & 500 &  21 & 299 &  1 & 567 &  1 & 768 &  0 & 031 & 1 & 375 & 0 & 158 \\
216437*     &   3 & 004 & 0 & 447 &  26 & 985 &  1 & 857 &  1 & 470 &  0 & 019 & 1 & 088 & 0 & 179 \\
216770      &   1 & 813 & 1 & 000 &  38 & 656 &  5 &  99 &  0 & 997 &  0 & 027 & 1 & 388 & 0 & 788 \\
217014      &   2 & 178 & 0 & 367 &  29 & 467 &  0 & 766 &  1 & 159 &  0 & 010 & 1 & 093 & 0 & 187 \\
219828      &   3 & 450 & 1 & 000 &  28 & 476 &  1 & 439 &  1 & 842 &  0 & 128 & 1 & 053 & 0 & 318 \\
222582      &   2 & 290 & 0 & 500 &  25 & 032 &  1 & 786 &  1 & 128 &  0 & 030 & 1 & 003 & 0 & 232 \\
224693*     &   3 & 799 & 0 & 447 &  29 & 735 &  1 & 487 &  1 & 831 &  0 & 148 & 1 & 218 & 0 & 184 \\
OGLE-TR-56  &   3 & 200 & 1 & 000 &  26 & 312 &  2 & 204 &  1 & 234 &  0 & 042 & 1 & 347 & 0 & 438 \\
OGLE-TR-111 &   5 & 000 & 1 & 000 &  37 & 964 &  6 & 118 &  0 & 829 &  0 & 020 & 4 & 518 & 1 & 165 \\

\label{tab:2}

\end{longtable}

\end{center}

\twocolumn

\onecolumn

\begin{center}

\begin{longtable}{lr@{.}lr@{.}lr@{.}lr@{.}lr@{.}lr@{.}lr@{.}lr@{.}lr@{.}lcc}
\caption[]{Adopted parameters and $\sin i$ estimates for extra-solar
planet host stars for which we have carried out a Markov-chain Monte
Carlo analysis. Extra-solar planets with $\sin i$ values more than
1-$\sigma$ greater than 1 (see Table~\ref{tab:2}) were excluded from
this analysis. Columns 1--7 are described in Table~\ref{tab:2}. For
stars with multiple planets, the first row gives the full planet name,
and subsequent planets are indicated in the following rows by their
designated letter only (e.g. `c', `d', etc.). Sub-giants are indicated
with an asterisk. Column  8 lists the calculated $\sin i$'s for each
star given the adopted $v \sin i,~P_{rot}$~and $R_*$, and columns 9 \&
10 list the two-tailed 1-$\sigma$ error bars on $\sin i$. Column 11
lists the exoplanet mass (in Jupiter masses) after applying the $\sin
i$ correction in column 8. Finally, column 12 gives the transit
probability for each extra-solar planet, where 1 indicates a 100\%
probability that the system shows transits.} \\

\hline \multicolumn{1}{c}{HD or} & \multicolumn{2}{c}{$v \sin i$} &
\multicolumn{2}{c}{$\sigma_v$} & \multicolumn{2}{c}{$P_{rot}$} &
\multicolumn{2}{c}{$\sigma_P$} & \multicolumn{2}{c}{$R_*$} &
\multicolumn{2}{c}{$\sigma_R$} & \multicolumn{2}{c}{$\sin i$} &
\multicolumn{2}{c}{$\sigma_-$} & \multicolumn{2}{c}{$\sigma_+$} & Mass
& prob. \\ \multicolumn{1}{c}{Alt. Name} & \multicolumn{2}{c}{(km
s$^{-1}$)} & \multicolumn{2}{c}{} & \multicolumn{2}{c}{(days)} &
\multicolumn{2}{c}{} & \multicolumn{2}{c}{($R_{\odot}$)} &
\multicolumn{2}{c}{} & \multicolumn{2}{c}{} & \multicolumn{2}{c}{} &
\multicolumn{2}{c}{} & ($M_J$) \\ \multicolumn{1}{c}{(1)} &
\multicolumn{2}{c}{(2)} & \multicolumn{2}{c}{(3)} &
\multicolumn{2}{c}{(4)} & \multicolumn{2}{c}{(5)} &
\multicolumn{2}{c}{(6)} & \multicolumn{2}{c}{(7)} &
\multicolumn{2}{c}{(8)} & \multicolumn{2}{c}{(9)} &
\multicolumn{2}{c}{(10)} & (11) & (12) \\ \hline \endfirsthead

\multicolumn{21}{c} {{\bfseries \tablename\ \thetable{} -- continued}}
\\ \hline

\multicolumn{1}{c}{Name or} & \multicolumn{2}{c}{$v \sin i$} &
\multicolumn{2}{c}{$\sigma_v$}& \multicolumn{2}{c}{$P_{rot}$} &
\multicolumn{2}{c}{$\sigma_P$} & \multicolumn{2}{c}{$R_*$} &
\multicolumn{2}{c}{$\sigma_R$} & \multicolumn{2}{c}{$\sin i$} &
\multicolumn{2}{c}{$\sigma_-$} & \multicolumn{2}{c}{$\sigma_+$} & Mass
& prob. \\

\multicolumn{1}{c}{Alt. Name} & \multicolumn{2}{c}{(km s$^{-1}$)} &
\multicolumn{2}{c}{} & \multicolumn{2}{c}{(days)} &
\multicolumn{2}{c}{} & \multicolumn{2}{c}{($R_{\odot}$)} &
\multicolumn{2}{c}{} & \multicolumn{2}{c}{} & \multicolumn{2}{c}{} &
\multicolumn{2}{c}{} & ($M_J$) \\

\multicolumn{1}{c}{(1)} & \multicolumn{2}{c}{(2)} &
\multicolumn{2}{c}{(3)} & \multicolumn{2}{c}{(4)} &
\multicolumn{2}{c}{(5)} & \multicolumn{2}{c}{(6)} &
\multicolumn{2}{c}{(7)} & \multicolumn{2}{c}{(8)} &
\multicolumn{2}{c}{(9)} & \multicolumn{2}{c}{(10)} & (11) & (12) \\
\hline \endhead

1237        &   5 & 510 & 1 & 000 &   4 & 314 &  3 & 213 &  0 & 901 &  0 & 014 & 0 & 527 & 0 & 365 & 0 & 413 &  6.295 & 0.004 \\
2638        &   1 & 100 & 1 & 000 &  38 & 832 &  4 & 626 &  0 & 926 &  0 & 057 & 0 & 905 & 0 & 153 & 0 & 094 &  0.530 & 0.055 \\
4203        &   1 & 229 & 0 & 500 &  43 & 015 &  1 & 550 &  1 & 403 &  0 & 075 & 0 & 744 & 0 & 211 & 0 & 213 &  2.215 & 0.002 \\
4308        &   0 & 400 & 0 & 447 &  22 & 524 &  2 &  12 &  1 & 029 &  0 & 009 & 0 & 174 & 0 & 172 & 0 & 306 &  0.269 & 0.000 \\
6434        &   2 & 149 & 1 & 000 &  17 & 235 &  1 & 620 &  0 & 910 &  0 & 031 & 0 & 811 & 0 & 176 & 0 & 166 &  0.591 & 0.013 \\
8574        &   4 & 327 & 0 & 386 &  17 & 073 &  0 & 884 &  1 & 383 &  0 & 029 & 0 & 999 & 0 & 049 & 0 & 000 &  2.230 & 0.009 \\
10647       &   5 & 464 & 0 & 447 &   7 & 669 &  1 &  38 &  1 & 101 &  0 & 014 & 0 & 756 & 0 & 155 & 0 & 160 &  1.203 & 0.001 \\
10697       &   2 & 479 & 0 & 500 &  34 & 273 &  1 & 181 &  1 & 791 &  0 & 032 & 0 & 941 & 0 & 069 & 0 & 058 &  6.502 & 0.002 \\
12661       &   1 & 300 & 0 & 500 &  37 & 253 &  2 & 457 &  1 & 145 &  0 & 025 & 0 & 834 & 0 & 142 & 0 & 151 &  2.755 & 0.003 \\
c & & & & & & & & & & & & & & & & & & &  1.881 & 0.001 \\
16141       &   1 & 743 & 0 & 447 &  31 & 839 &  1 & 554 &  1 & 453 &  0 & 043 & 0 & 754 & 0 & 187 & 0 & 194 &  0.305 & 0.006 \\
17051       &   5 & 599 & 0 & 304 &   7 & 921 &  1 & 626 &  1 & 156 &  0 & 012 & 0 & 756 & 0 & 180 & 0 & 191 &  2.565 & 0.002 \\
17156       &   2 & 600 & 0 & 500 &  22 & 138 &  1 & 118 &  1 & 504 &  0 & 056 & 0 & 754 & 0 & 165 & 0 & 177 &  4.123 & 0.011 \\
19994       &   8 & 511 & 0 & 408 &  10 & 783 &  1 & 682 &  1 & 698 &  0 & 028 & 0 & 999 & 0 & 066 & 0 & 000 &  2.001 & 0.006 \\
20367       &   3 & 290 & 1 & 000 &   5 & 465 &  1 & 497 &  1 & 200 &  0 & 025 & 0 & 293 & 0 & 135 & 0 & 184 &  3.650 & 0.000 \\
20782       &   2 & 391 & 0 & 447 &  20 & 462 &  1 & 980 &  1 & 124 &  0 & 021 & 0 & 864 & 0 & 118 & 0 & 118 &  2.083 & 0.002 \\
22049       &   1 & 898 & 0 & 257 &  11 & 679 &  6 & 801 &  0 & 721 &  0 & 007 & 0 & 588 & 0 & 295 & 0 & 359 &  2.632 & 0.001 \\
23079       &   2 & 990 & 0 & 500 &  17 & 117 &  1 & 182 &  1 & 128 &  0 & 014 & 0 & 895 & 0 & 091 & 0 & 094 &  2.914 & 0.002 \\
23596       &   3 & 956 & 0 & 381 &  21 & 251 &  1 & 108 &  1 & 583 &  0 & 048 & 0 & 999 & 0 & 055 & 0 & 000 &  7.191 & 0.003 \\
27442*      &   2 & 873 & 0 & 257 &  89 & 184 & 15 & 674 &  4 & 335 &  0 & 427 & 0 & 999 & 0 & 080 & 0 & 000 &  1.280 & 0.010 \\
33564       &  12 & 390 & 0 & 937 &   6 & 802 &  0 & 429 &  1 & 503 &  0 & 024 & 0 & 999 & 0 & 037 & 0 & 000 &  9.100 & 0.008 \\
33636       &   3 & 080 & 0 & 500 &  16 & 697 &  0 & 966 &  1 & 003 &  0 & 022 & 0 & 999 & 0 & 094 & 0 & 000 &  9.282 & 0.001 \\
37124       &   1 & 219 & 0 & 500 &  27 & 311 &  0 & 650 &  1 & 006 &  0 & 027 & 0 & 645 & 0 & 258 & 0 & 278 &  0.946 & 0.003 \\
c & & & & & & & & & & & & & & & & & & &  1.059 & 0.000 \\
d & & & & & & & & & & & & & & & & & & &  0.930 & 0.001 \\
38529       &   3 & 899 & 0 & 500 &  37 & 761 &  2 & 210 &  2 & 750 &  0 & 079 & 0 & 999 & 0 & 068 & 0 & 000 &  0.780 & 0.079 \\
c & & & & & & & & & & & & & & & & & & & 12.705 & 0.002 \\
39091       &   3 & 140 & 0 & 500 &  17 & 328 &  1 & 583 &  1 & 161 &  0 & 010 & 0 & 924 & 0 & 067 & 0 & 074 & 11.193 & 0.001 \\
40979       &   7 & 429 & 0 & 500 &   7 & 896 &  0 & 948 &  1 & 205 &  0 & 020 & 0 & 964 & 0 & 047 & 0 & 035 &  3.443 & 0.005 \\
41004A      &   1 & 609 & 1 & 000 &  26 & 897 &  6 & 627 &  1 & 016 &  0 & 045 & 0 & 833 & 0 & 136 & 0 & 160 &  2.759 & 0.001 \\
45350       &   1 & 370 & 0 & 500 &  39 & 402 &  1 & 921 &  1 & 299 &  0 & 035 & 0 & 818 & 0 & 155 & 0 & 160 &  2.187 & 0.001 \\
46375       &   0 & 859 & 0 & 500 &  43 & 876 &  3 & 514 &  1 & 024 &  0 & 027 & 0 & 738 & 0 & 233 & 0 & 223 &  0.337 & 0.058 \\
49674       &   0 & 419 & 0 & 500 &  27 & 226 &  1 & 740 &  0 & 974 &  0 & 025 & 0 & 241 & 0 & 239 & 0 & 467 &  0.477 & 0.007 \\
50554       &   3 & 675 & 0 & 356 &  14 & 665 &  0 & 474 &  1 & 134 &  0 & 021 & 0 & 939 & 0 & 056 & 0 & 056 &  5.217 & 0.002 \\
62509       &   1 & 331 & 0 & 668 & 135 & 000 & 13 & 500 &  8 & 738 &  0 & 098 & 0 & 392 & 0 & 329 & 0 & 523 &  7.393 & 0.004 \\
69830       &   0 & 700 & 0 & 353 &  36 & 452 &  1 & 929 &  0 & 892 &  0 & 010 & 0 & 554 & 0 & 311 & 0 & 339 &  0.060 & 0.015 \\
c & & & & & & & & & & & & & & & & & & &  0.069 & 0.006 \\
d & & & & & & & & & & & & & & & & & & &  0.105 & 0.002 \\
70573       &  12 & 300 & 1 & 000 &   3 & 295 & 31 & 226 &  0 & 846 &  0 & 251 & 0 & 991 & 0 & 244 & 0 & 000 &  6.155 & 0.002 \\
70642       &   0 & 299 & 0 & 500 &  28 & 829 &  3 & 276 &  1 & 016 &  0 & 013 & 0 & 167 & 0 & 166 & 0 & 520 & 11.932 & 0.000 \\
72659       &   2 & 209 & 0 & 500 &  20 & 731 &  0 & 759 &  1 & 458 &  0 & 043 & 0 & 616 & 0 & 177 & 0 & 191 &  4.798 & 0.000 \\
73256       &   3 & 260 & 1 & 000 &  13 & 912 &  3 & 897 &  0 & 966 &  0 & 025 & 0 & 921 & 0 & 113 & 0 & 078 &  2.029 & 0.077 \\
73526       &   2 & 620 & 0 & 500 &  35 & 643 &  2 & 433 &  1 & 505 &  0 & 077 & 0 & 999 & 0 & 083 & 0 & 000 &  2.900 & 0.009 \\
c & & & & & & & & & & & & & & & & & & &  2.500 & 0.005 \\
74156       &   4 & 217 & 0 & 389 &  18 & 202 &  0 & 891 &  1 & 627 &  0 & 063 & 0 & 935 & 0 & 061 & 0 & 059 &  2.011 & 0.016 \\
c & & & & & & & & & & & & & & & & & & &  8.588 & 0.001 \\
d & & & & & & & & & & & & & & & & & & &  0.424 & 0.004 \\
75289       &   4 & 139 & 0 & 500 &  16 & 839 &  1 & 201 &  1 & 271 &  0 & 016 & 0 & 999 & 0 & 063 & 0 & 000 &  0.410 & 0.148 \\
76700       &   1 & 350 & 0 & 500 &  36 & 599 &  4 & 721 &  1 & 372 &  0 & 031 & 0 & 709 & 0 & 215 & 0 & 238 &  0.277 & 0.049 \\
80606       &   1 & 431 & 0 & 384 &  42 & 254 &  2 &  62 &  0 & 941 &  0 & 192 & 0 & 999 & 0 & 128 & 0 & 000 &  3.410 & 0.008 \\
81040       &   2 & 000 & 1 & 000 &   9 & 085 &  2 &  12 &  0 & 887 &  0 & 033 & 0 & 400 & 0 & 290 & 0 & 368 & 17.136 & 0.000 \\
82943       &   1 & 420 & 0 & 447 &  21 & 892 &  1 & 912 &  1 & 098 &  0 & 018 & 0 & 566 & 0 & 250 & 0 & 254 &  3.090 & 0.001 \\
c & & & & & & & & & & & & & & & & & & &  3.550 & 0.001 \\
83443       &   1 & 303 & 0 & 447 &  35 & 999 &  4 &  37 &  1 & 058 &  0 & 024 & 0 & 868 & 0 & 105 & 0 & 127 &  0.460 & 0.069 \\
88133*      &   2 & 185 & 0 & 353 &  49 & 838 &  3 & 263 &  2 & 080 &  0 & 113 & 0 & 999 & 0 & 093 & 0 & 000 &  0.220 & 0.166 \\
89307       &   2 & 879 & 0 & 500 &  17 & 155 &  1 & 369 &  1 & 075 &  0 & 024 & 0 & 909 & 0 & 086 & 0 & 084 &  3.001 & 0.001 \\
89744       &   9 & 208 & 0 & 447 &   9 & 000 &  6 & 785 &  2 & 126 &  0 & 041 & 0 & 763 & 0 & 191 & 0 & 212 & 10.461 & 0.006 \\
92788       &   0 & 567 & 0 & 447 &  33 & 611 &  1 & 702 &  1 & 049 &  0 & 020 & 0 & 375 & 0 & 369 & 0 & 427 & 10.276 & 0.001 \\
93083       &   0 & 900 & 1 & 000 &  48 & 549 &  3 & 434 &  0 & 874 &  0 & 026 & 0 & 990 & 0 & 233 & 0 & 009 &  0.374 & 0.004 \\
95128       &   2 & 830 & 0 & 231 &  21 & 113 &  0 & 373 &  1 & 220 &  0 & 012 & 0 & 969 & 0 & 033 & 0 & 030 &  2.681 & 0.002 \\
c & & & & & & & & & & & & & & & & & & &  0.474 & 0.002 \\
101930      &   0 & 699 & 1 & 000 &  46 & 575 &  3 & 485 &  0 & 907 &  0 & 028 & 0 & 710 & 0 & 238 & 0 & 253 &  0.422 & 0.007 \\
102117      &   1 & 004 & 0 & 447 &  37 & 555 &  1 & 342 &  1 & 314 &  0 & 026 & 0 & 580 & 0 & 320 & 0 & 298 &  0.296 & 0.009 \\
104985      &   2 & 699 & 1 & 100 & 120 & 982 & 29 & 627 & 10 & 273 &  1 & 176 & 0 & 635 & 0 & 256 & 0 & 299 &  9.913 & 0.010 \\
106252      &   1 & 778 & 0 & 223 &  20 & 523 &  0 & 812 &  1 & 107 &  0 & 022 & 0 & 647 & 0 & 092 & 0 & 102 & 10.517 & 0.000 \\
107148      &   0 & 705 & 0 & 447 &  32 & 451 &  1 & 660 &  1 & 186 &  0 & 034 & 0 & 385 & 0 & 369 & 0 & 371 &  0.545 & 0.002 \\
108147      &   5 & 939 & 0 & 447 &   8 & 867 &  1 & 267 &  1 & 220 &  0 & 019 & 0 & 855 & 0 & 117 & 0 & 117 &  0.468 & 0.032 \\
109749      &   2 & 399 & 0 & 447 &  27 & 091 &  1 & 810 &  1 & 243 &  0 & 075 & 0 & 999 & 0 & 103 & 0 & 000 &  0.280 & 0.077 \\
114386      &   0 & 589 & 0 & 500 &  35 & 568 &  3 & 658 &  0 & 778 &  0 & 021 & 0 & 545 & 0 & 380 & 0 & 375 &  1.815 & 0.001 \\
114729      &   2 & 290 & 0 & 500 &  18 & 836 &  0 & 333 &  1 & 439 &  0 & 029 & 0 & 590 & 0 & 161 & 0 & 166 &  1.389 & 0.000 \\
114783      &   0 & 869 & 0 & 500 &  45 & 202 &  2 & 447 &  0 & 807 &  0 & 011 & 0 & 968 & 0 & 181 & 0 & 031 &  1.022 & 0.002 \\
117176      &   2 & 827 & 0 & 249 &  35 & 463 &  3 &   4 &  1 & 825 &  0 & 025 & 0 & 999 & 0 & 052 & 0 & 000 &  7.440 & 0.017 \\
117207      &   1 & 050 & 0 & 500 &  37 & 238 &  1 & 296 &  1 & 128 &  0 & 024 & 0 & 687 & 0 & 260 & 0 & 259 &  2.997 & 0.001 \\
120136      &  14 & 735 & 0 & 173 &   4 & 000 & 0 & 400 &  1 & 426 &  0 & 016 & 0 & 777 & 0 & 020 & 0 & 221 &  5.015 & 0.109 \\
121504      &   3 & 299 & 1 & 000 &  11 & 397 &  2 & 162 &  1 & 196 &  0 & 045 & 0 & 626 & 0 & 245 & 0 & 282 &  1.421 & 0.004 \\
125612      &   2 & 680 & 0 & 447 &  17 & 625 &  0 & 965 &  1 & 030 &  0 & 040 & 0 & 911 & 0 & 091 & 0 & 082 &  3.510 & 0.002 \\
128311      &   3 & 649 & 0 & 500 &  10 & 778 &  2 & 714 &  0 & 769 &  0 & 011 & 0 & 999 & 0 & 127 & 0 & 000 &  2.180 & 0.002 \\
c & & & & & & & & & & & & & & & & & & &  3.210 & 0.002 \\
134987      &   2 & 169 & 0 & 500 &  33 & 778 &  1 & 649 &  1 & 225 &  0 & 018 & 0 & 999 & 0 & 106 & 0 & 000 &  1.580 & 0.005 \\
136118      &   7 & 330 & 0 & 500 &   9 & 845 &  1 & 148 &  1 & 744 &  0 & 044 & 0 & 818 & 0 & 125 & 0 & 131 & 14.536 & 0.001 \\
141937      &   1 & 923 & 0 & 447 &  15 & 533 &  2 & 300 &  1 & 079 &  0 & 036 & 0 & 550 & 0 & 201 & 0 & 233 & 17.636 & 0.000 \\
142415      &   3 & 403 & 0 & 447 &  12 & 344 &  2 & 567 &  1 & 039 &  0 & 022 & 0 & 806 & 0 & 159 & 0 & 158 &  2.010 & 0.002 \\
143761      &   1 & 420 & 0 & 300 &  17 & 000 &  7 & 223 &  1 & 328 &  0 & 015 & 0 & 350 & 0 & 232 & 0 & 331 &  2.969 & 0.001 \\
145675      &   1 & 560 & 0 & 500 &  48 & 500 &  1 & 137 &  0 & 984 &  0 & 009 & 0 & 999 & 0 & 138 & 0 & 000 &  4.640 & 0.001 \\
147506      &  19 & 800 & 1 & 600 &   4 & 045 &  0 & 373 &  1 & 600 &  0 & 117 & 0 & 989 & 0 & 069 & 0 & 010 &  8.709 & 0.104 \\
147513      &   1 & 475 & 0 & 353 &   8 & 525 &  2 & 233 &  0 & 947 &  0 & 011 & 0 & 259 & 0 & 099 & 0 & 130 &  3.849 & 0.000 \\
150706      &   3 & 650 & 0 & 325 &   9 & 428 &  2 & 195 &  0 & 959 &  0 & 012 & 0 & 691 & 0 & 191 & 0 & 242 &  1.446 & 0.002 \\
154857      &   1 & 439 & 0 & 500 &  31 & 520 &  2 & 162 &  2 & 466 &  0 & 101 & 0 & 358 & 0 & 140 & 0 & 157 &  5.023 & 0.000 \\
159868      &   2 & 100 & 0 & 500 &  35 & 537 &  2 & 426 &  1 & 888 &  0 & 072 & 0 & 775 & 0 & 165 & 0 & 181 &  2.191 & 0.001 \\
162020      &   2 & 235 & 0 & 447 &   1 & 620 &  1 &  27 &  0 & 746 &  0 & 025 & 0 & 093 & 0 & 080 & 0 & 112 & 147.849 & 0.000 \\
168443      &   2 & 100 & 0 & 447 &  38 & 606 &  0 & 675 &  1 & 595 &  0 & 030 & 0 & 999 & 0 & 117 & 0 & 000 &  8.021 & 0.016 \\
c & & & & & & & & & & & & & & & & & & & 18.101 & 0.002 \\
168746      &   0 & 500 & 0 & 408 &  34 & 774 &  1 & 738 &  1 & 132 &  0 & 027 & 0 & 294 & 0 & 287 & 0 & 391 &  0.780 & 0.004 \\
169830      &   3 & 724 & 0 & 447 &   9 & 625 &  1 & 810 &  1 & 838 &  0 & 036 & 0 & 384 & 0 & 089 & 0 & 102 &  7.497 & 0.000 \\
c & & & & & & & & & & & & & & & & & & & 10.517 & 0.000 \\
170469      &   1 & 699 & 0 & 500 &  31 & 518 &  1 &  86 &  1 & 302 &  0 & 042 & 0 & 821 & 0 & 165 & 0 & 152 &  0.815 & 0.001 \\
175541      &   2 & 899 & 0 & 500 &  58 & 171 &  1 & 324 &  3 & 800 &  0 & 008 & 0 & 880 & 0 & 109 & 0 & 102 &  0.693 & 0.006 \\
177830*     &   2 & 540 & 0 & 500 &  65 & 711 &  6 & 921 &  3 & 129 &  0 & 105 & 0 & 999 & 0 & 111 & 0 & 000 &  1.280 & 0.006 \\
178911B     &   1 & 939 & 0 & 500 &  36 & 250 &  2 &  24 &  1 & 130 &  0 & 183 & 0 & 999 & 0 & 126 & 0 & 000 &  6.294 & 0.012 \\
179949      &   7 & 019 & 0 & 500 &   7 & 700 &  0 & 486 &  1 & 227 &  0 & 020 & 0 & 868 & 0 & 088 & 0 & 097 &  1.094 & 0.065 \\
183263      &   1 & 560 & 0 & 500 &  28 & 001 &  1 & 367 &  1 & 236 &  0 & 046 & 0 & 695 & 0 & 222 & 0 & 235 &  5.302 & 0.001 \\
186427      &   2 & 253 & 0 & 315 &  29 & 343 &  0 & 767 &  1 & 167 &  0 & 009 & 0 & 999 & 0 & 063 & 0 & 000 &  1.680 & 0.003 \\
187085      &   5 & 099 & 0 & 500 &  14 & 349 &  1 & 206 &  1 & 331 &  0 & 045 & 0 & 999 & 0 & 057 & 0 & 000 &  0.750 & 0.003 \\
187123      &   2 & 149 & 0 & 500 &  26 & 804 &  1 & 375 &  1 & 185 &  0 & 023 & 0 & 962 & 0 & 098 & 0 & 037 &  0.540 & 0.095 \\
c & & & & & & & & & & & & & & & & & & &  2.025 & 0.001 \\
189733      &   2 & 730 & 0 & 832 &  13 & 230 &  5 & 338 &  0 & 760 &  0 & 011 & 0 & 938 & 0 & 131 & 0 & 061 &  1.226 & 0.075 \\
190228      &   1 & 850 & 0 & 500 &  47 & 970 &  3 & 216 &  2 & 473 &  0 & 083 & 0 & 708 & 0 & 211 & 0 & 221 &  7.047 & 0.001 \\
190647      &   1 & 969 & 0 & 832 &  40 & 977 &  1 & 410 &  1 & 496 &  0 & 061 & 0 & 999 & 0 & 180 & 0 & 000 &  1.900 & 0.002 \\
192699      &   1 & 899 & 0 & 500 &  59 & 813 &  2 & 888 &  3 & 923 &  0 & 058 & 0 & 574 & 0 & 186 & 0 & 185 &  4.351 & 0.000 \\
195019      &   2 & 470 & 0 & 500 &  29 & 074 &  2 & 488 &  1 & 464 &  0 & 035 & 0 & 969 & 0 & 096 & 0 & 030 &  3.818 & 0.029 \\
196050      &   3 & 235 & 0 & 447 &  23 & 282 &  7 & 293 &  1 & 321 &  0 & 039 & 0 & 999 & 0 & 120 & 0 & 000 &  3.001 & 0.002 \\
196885*     &   7 & 750 & 0 & 500 &  12 & 306 &  0 & 672 &  1 & 387 &  0 & 027 & 0 & 999 & 0 & 017 & 0 & 000 &  2.960 & 0.005 \\
202206      &   2 & 299 & 0 & 500 &  22 & 980 &  3 & 585 &  1 & 064 &  0 & 032 & 0 & 984 & 0 & 125 & 0 & 015 & 17.674 & 0.004 \\
c & & & & & & & & & & & & & & & & & & &  2.478 & 0.001 \\
208487      &   4 & 610 & 0 & 500 &  12 & 412 &  1 & 134 &  1 & 150 &  0 & 035 & 0 & 983 & 0 & 069 & 0 & 016 &  0.457 & 0.009 \\
209458      &   4 & 280 & 0 & 367 &  14 & 914 &  0 & 629 &  1 & 150 &  0 & 029 & 0 & 999 & 0 & 040 & 0 & 000 &  0.690 & 0.186 \\
210702      &   1 & 699 & 0 & 500 &  69 & 061 &  5 & 140 &  4 & 449 &  0 & 069 & 0 & 527 & 0 & 216 & 0 & 215 &  3.793 & 0.000 \\
212301      &   6 & 220 & 1 & 000 &  11 & 340 &  0 & 492 &  1 & 172 &  0 & 030 & 0 & 999 & 0 & 067 & 0 & 000 &  0.450 & 0.182 \\
213240      &   3 & 969 & 0 & 609 &  17 & 022 &  4 & 748 &  1 & 536 &  0 & 036 & 0 & 867 & 0 & 110 & 0 & 122 &  5.185 & 0.002 \\
216437*     &   3 & 004 & 0 & 447 &  26 & 985 &  1 & 857 &  1 & 470 &  0 & 019 & 0 & 999 & 0 & 077 & 0 & 000 &  2.100 & 0.002 \\
217014      &   2 & 178 & 0 & 367 &  29 & 467 &  0 & 766 &  1 & 159 &  0 & 010 & 0 & 999 & 0 & 081 & 0 & 000 &  0.468 & 0.106 \\
219828      &   3 & 450 & 1 & 000 &  28 & 476 &  1 & 439 &  1 & 842 &  0 & 128 & 0 & 999 & 0 & 145 & 0 & 000 &  0.066 & 0.110 \\
221287      &   5 & 607 & 0 & 832 &   4 & 586 &  0 & 424 &  1 & 126 &  0 & 033 & 0 & 452 & 0 & 085 & 0 & 090 &  6.830 & 0.000 \\
222404      &   1 & 500 & 1 & 000 &  68 & 020 &  4 & 626 &  4 & 511 &  0 & 527 & 0 & 454 & 0 & 394 & 0 & 411 &  3.522 & 0.001 \\
222582      &   2 & 290 & 0 & 500 &  25 & 032 &  1 & 786 &  1 & 128 &  0 & 030 & 0 & 996 & 0 & 119 & 0 & 003 &  5.129 & 0.003 \\
231701      &   4 & 000 & 0 & 500 &  10 & 276 &  0 & 383 &  1 & 372 &  0 & 131 & 0 & 594 & 0 & 112 & 0 & 124 &  2.995 & 0.000 \\
330075      &   0 & 699 & 0 & 200 &  47 & 365 &  3 & 209 &  1 & 008 &  0 & 062 & 0 & 649 & 0 & 231 & 0 & 250 &  1.171 & 0.028 \\
TrES-1      &   1 & 195 & 0 & 169 &  33 & 528 &  5 & 968 &  0 & 824 &  0 & 015 & 0 & 970 & 0 & 091 & 0 & 029 &  0.629 & 0.081 \\
TrES-2      &   2 & 000 & 1 & 000 &  24 & 783 &  1 & 622 &  1 & 000 &  0 & 035 & 0 & 986 & 0 & 188 & 0 & 013 &  1.214 & 0.081 \\
HAT-P-1     &   2 & 200 & 0 & 200 &  19 & 711 &  1 &  44 &  1 & 149 &  0 & 100 & 0 & 747 & 0 & 123 & 0 & 141 &  0.701 & 0.014 \\
OGLE-TR-10  &   3 & 000 & 2 & 000 &  15 & 836 &  1 & 925 &  1 & 143 &  0 & 042 & 0 & 809 & 0 & 154 & 0 & 178 &  0.778 & 0.068 \\
OGLE-TR-56  &   3 & 200 & 1 & 000 &  26 & 312 &  2 & 204 &  1 & 234 &  0 & 042 & 0 & 999 & 0 & 136 & 0 & 000 &  1.290 & 0.218 \\
OGLE-TR-113 &   9 & 000 & 3 & 000 &   3 & 244 &  0 & 748 &  0 & 765 &  0 & 025 & 0 & 763 & 0 & 191 & 0 & 201 &  1.730 & 0.082 \\

\label{tab:3}

\end{longtable}

\end{center}

\twocolumn

\begin{table}
\caption[]{Summary of known transiting planets. All the listed $\sin
i$'s are from the Markov-chain Monte Carlo analysis, except
OGLE-TR-111b which was not included in the MCMC analysis on account of
its discrepant $\sin i$ value. The two-tailed 1-$\sigma$ error bars on
$\sin i$, $\sigma_-$ and $\sigma_+$ are also listed. The final column
indicates the transit probability as calculated from the MCMC analysis,
where available.}

\begin{tabular}{lr@{.}lr@{.}lr@{.}lc} \hline
Name & \multicolumn{2}{c}{$\sin i$} & \multicolumn{2}{c}{$\sigma_-$} &
\multicolumn{2}{c}{$\sigma_+$} & Trans. prob. \\
\hline
HAT-P-1b      &   0&747 & 0&123 & 0&141 & 0.014 \\
HAT-P-2b      &   0&989 & 0&069 & 0&010 & 0.104 \\
HD17156       &   0&754 & 0&165 & 0&177 & 0.011 \\ 
HD189733b     &   0&933 & 0&131 & 0&061 & 0.075 \\
HD209458b     &   0&999 & 0&040 & 0&000 & 0.186 \\
OGLE-TR-10b   &   0&809 & 0&154 & 0&178 & 0.068 \\
OGLE-TR-56b   &   0&999 & 0&134 & 0&000 & 0.219 \\
OGLE-TR-111b  &   4&518 & 1&165 & 1&165 & ...   \\
OGLE-TR-113b  &   0&763 & 0&191 & 0&201 & 0.082 \\
TrES-1b       &   0&970 & 0&091 & 0&029 & 0.081 \\
TrES-2b       &   0&986 & 0&188 & 0&013 & 0.081 \\

\label{tab:trans}

\end{tabular}
\end{table}

\begin{table}
\caption[]{The 20 most probable transiting extra-solar planets as
determined from spectroscopic data and the MCMC analysis. Only
spectroscopically-discovered planets are included in this
table. Column 1 gives the common name for the extra-solar planet,
column 2 the extra-solar planet's orbital period in days, and column 3
the transit probability as determined from the MCMC analysis}

\begin{tabular}{lr@{.}lc} \hline
Name & \multicolumn{2}{c} {$P_{orb}$ (days)} & Trans. prob. \\ \hline

HD 212301b     &   2&457      & 0.182 \\
HD 88133b      &   3&41       & 0.166 \\
HD 75289b      &   3&51       & 0.148 \\
HD 219828b     &   3&8335     & 0.110 \\
$\tau$~Boo-b   &   3&3135     & 0.109 \\
51 Peg         &   4&23077    & 0.106 \\
HD 187123b     &   3&097      & 0.095 \\
HD 38529b      &  14&309      & 0.079 \\
CS Pyx-b       &   2&54858    & 0.077 \\
HD 109749b     &   5&24       & 0.077 \\
HD 83443b      &   2&985625   & 0.069 \\
HD 179949b     &   3&0925     & 0.065 \\
HD 46375b      &   3&024      & 0.058 \\
HD 2638b       &   3&4442     & 0.055 \\
HD 76700b      &   3&971      & 0.049 \\
HD 108147b     &  10&901      & 0.032 \\
HD 195019      &  18&20163    & 0.029 \\
HD 330075b     &   3&369      & 0.028 \\
HD 117176      & 116&689      & 0.017 \\
HD 74156       &  51&65       & 0.016 \\

\label{tab:prob}

\end{tabular}
\end{table}

\section*{\sc Acknowledgments}

Much of this work was carried out while CAW was supported by a PPARC/STFC
Postdoctoral Fellowship. SPL acknowledges the support of an RCUK fellowship.
The authors would also like to thank the anonymous referee whose detailed
and valuable comments helped substantially improve the quality of this paper.

\section*{appendix A: Notes on specific systems with sin i greater than 1}

From Tables~\ref{tab:2} and~\ref{tab:3} we can see that out of a total
of 154 extra-solar planet hosts with sufficient data, 119 (77 per
cent) yield $\sin i <$ 1 or are within {1-$\sigma$} of $\sin i =
1$. In this Section we discuss why some systems have calculated $\sin
i$'s significantly (i.e. more than 1-$\sigma$) greater than 1.

\subsection*{Sub-giants}

Of the 35 extra-solar planet host stars with $\sin i$ significantly
greater than 1, ten are classified as sub-giants. Since the
\cite{noyes84} chromospheric index -- rotation rate relationship is
calibrated for main-sequence stars only, we believe that the rotation
periods of these stars determined from $R'_{HK}$ measurements may be
incorrect. We have indicated the sub-giants with asterisks in
Table~\ref{tab:2}. Other systems where we can highlight potential
problems which may result in values of $\sin i >$ 1 are discussed
briefly below.

\subsection{HD 142}

In addition to being classified as a sub-giant, the ($B-V$) colour
of this star may be contaminated by a nearby companion as reported by
the NStED database.

\subsection*{HD 11506}

We note that \cite{fischer07} quote the stellar rotation period
determined from the $\log R'_{HK}$ measurements is 12.6 days. We, however,
derive a longer rotation period of 18.3 days using the $\log R'_{HK}$
value reported by \cite{fischer07} and the relationship from
\cite{noyes84}. We therefore believe the rotation period quoted by
\cite{fischer07} has been calculated incorrectly.

\subsection*{HD 13445}

There seems to be some confusion over the $v \sin i$ value for this
star. \cite{fischer05} quote 2.37 km s$^{-1}$, while \cite{saar97}
place an upper limit of 0.7 km s$^{-1}$. Assuming a $v \sin i$ of 2.37
km s$^{-1}$ results in a large $\sin i \sim $ 1.8, while adopting the
limit of 0.7 km s$^{-1}$ gives $\sin i \sim $ 0.5. Given the doubt
over $v \sin i$ for this star we have deemed this measurement to be
suspect.

\subsection*{HD 27442}

There is considerable doubt over the radius of this star, with
estimates ranging from 3.48--6.60 $R_{\odot}$. Furthermore, this star
is classified as a sub-giant, hence the rotation period derived from
the $\log R'_{HK}$ measurements is also likely to be inaccurate.

\subsection*{HD 27894}

This has an uncertain $v \sin i$, with only an upper limit of 1.5 km
s$^{-1}$ from \cite{moutou05}.

\subsection*{HD 28185}

This has an uncertain $v \sin i$, with estimates ranging from 1.82 --
3.00 km s$^{-1}$. While $v \sin i$ = 1.82 km s$^{-1}$ gives $\sin i$ =
1, we feel there is too much uncertainty in the $v \sin i$ values, and
hence have taken a weighted mean, placing HD 28185 in the $\sin i$
significantly greater than 1 category.

\subsection*{HD 75732}

The calculated rotational period of the star from $\log R'_{HK}$
measurements (42--47 days) is possibly related to the orbit of one of
its planets, 55 Cnc c, which has a measured orbital period of 43.93
days (see \citealt{marcy02}).

\subsection*{HD 86081}

We note that \cite{johnson06a} derived a stellar rotation period of
40.1 days from their measured $\log R'_{HK}$ using the calibration of
\cite{noyes84}. Employing the same $B-V$ colour and $\log R'_{HK}$
quoted by \cite{johnson06a} we determine a far shorter rotation period
of 27.7 days via the same relationship, and 24.83 days if we take the
value of $B-V$ = 0.641 from the NStED database. We therefore conclude
that the rotation period derived by \cite{johnson06a} is incorrect but,
despite the shorter rotation period we have calculated, we still
determine $\sin i$ = 1.590.

\subsection*{HD 145675}

While \cite{fischer05} quote a $v \sin i$ = 1.56 km s$^{-1}$,
\cite{naef04} quote an upper limit of $v \sin i <$ 1 km s$^{-1}$. This
upper limit would yield $\sin i < $0.832.  Given the apparent doubt
over $v \sin i$ we have decided to place this object in the $\sin i$
significantly greater than 1 category.

\subsection*{HD 216435}

\cite{jones03} have noted a discrepancy between the assigned
spectral-type of HD 216435 in the literature, which is either quoted as
G0V or G3IV. They find that HD 216435 lies 1 magnitude above the main
sequence, and hence this star is most likely a sub-giant.  The
rotation period determined from $\log R'_{HK}$ is therefore suspect.

\subsection*{OGLE-TR-111}
\label{sec:ogles}

While the OGLE extra-solar planets are all transiting systems,
published data for OGLE-TR-111 yields a $\sin i = 4.518 \pm 0.486$ and
is one of the most discrepant systems found in this work. We believe
that this is undoubtedly due to the faintness of the OGLE targets (all
OGLE extra-solar planet hosts have $I > 14$ whereas most extra-solar
planet hosts typically have $V-$band magnitudes around 8 -- 9), which
means that accurate spectroscopic follow-up is difficult.  In
addition, none of these systems have a long baseline of $R'_{HK}$
measurements, which means that the rotation periods are also not well
known. For these reasons, we believe that systematic errors in
one or more of the measurements have contributed
to the highly discrepant $\sin i$ obtained for OGLR-TR-111.

\subsection*{Summary}

In total, we can find plausible reasons explaining why 18 of the
extra-solar planet host stars yield $\sin i$'s significantly greater than
1. This still leaves 17 systems for which no explanation can be given
for their high $\sin i$ values.

\section*{Appendix B: Notes on specific systems with sin i less than 1}

In this Section we highlight any published data on stars that
appears incorrect, and justify any decisions that have been made
regarding the rejection of any published parameters from our
analysis. Any other special cases that apply are also indicated here,
such as the use of actual observed stellar rotation rates from
photometry instead of rotation rates derived from $\log R'_{HK}$
measurements, for example.

\subsection*{HD 1237}

The value of $P_{rot}$ = 12.6 days quoted on the Geneva Observatory
web-page and apparently derived from the \cite{noyes84} relationship
appears to be wrong. Using the Geneva Observatory's values of $\log
R'_{HK}$ = -4.27 and $B-V$ = 0.749, we derive $P_{rot}$ = 4.01 days. We
note that \cite{sbarnes01} use the same $B-V$ value, but a weaker
chromospheric activity index of $\log R'{HK}$ = -4.44 and derive a
rotation period of 10.4 days. Using the values of \cite{sbarnes01}, we
also derive 10.4 days, and thus conclude that the Geneva $P_{rot}$ is
quoted incorrectly.

\subsection*{HD 6434}

The Extrasolar Planets Encyclopedia quote the radius of HD 6434 as
0.57 $R_{\odot}$ (from \citealt{fracassini01}) and its spectral type
as G3 IV. The NStED database quotes the spectral type as G2--3
V. Given the spectral type, we find it highly unlikely that the radius
is actually 0.57 $R_{\odot}$, and instead use the value of 1.0
$R_{\odot}$ from the NStED database.

\subsection*{HD 16141}

The Extrasolar Planets Encyclopedia quotes a radius for HD 16141 of
1$R_{\odot}$ but provides no reference for this figure. Given this,
and that the radius is discrepant from other estimates obtained from
the literature (1.4 and 1.52 $R_{\odot}$), we have rejected this
radius estimate from Table~\ref{tab:1}.

\subsection*{HD 142022A}

\cite{eggenberger06} determined an upper limit of 48 days to the
rotation period of HD 142022A by combining their measured $v \sin i$
with the radius of the star estimated from evolutionary
models. Despite measuring $\log R'_{HK}$, they did not calculate the
rotation period using the \cite{noyes84} relationship. Using
\cite{eggenberger06}'s values for $\log R'_{HK}$ = -4.97 and $B-V$ =
0.790 we determine a rotation period of 39 days.

\subsection*{HD 170469}

\cite{fischer07} quote a $\log R'_{HK}$ = -5.06 and determine the
rotation period to be 13 days. Using the same value of $\log R'_{HK}$,
we determine the rotation period to be 30 days. Our period agrees
closely with that of \cite{wright04}, who find a rotation period of 31 days
from a very similar measurement of $\log R'_{HK}$ = -5.09. We
therefore assume that \cite{fischer07} have calculated the rotation
period incorrectly.

\subsection*{HD 217014}

The rotation period of 21.9 days has been used since this is a
measured rotation period from variability in the light curve, rather
than one estimated from $\log R'_{HK}$.

\onecolumn

\begin{center}
\begin{longtable}{lccll}

\caption[]{Compilation of chromospheric indices ($\log R'_{HK}$) for the stars
in Table 1. The spectral type of the host star is given in column 2. Entries
in bold give the grade assigned to each star (P = Poor, O = O.K., G = Good, and
E = Excellent) followed by the weighted mean of the $\log R'_{HK}$ measurements
and adopted error bar (see section~\ref{sec:rhk_data} for details). Reference numbers
are identical to those used in Table~\ref{tab:1}.}

\\

\hline 
\hline
Name & Type & $\log$ R'$_{HK}$ & Observations & Ref. \\
\hline
\endfirsthead

\multicolumn{5}{c} {{\bfseries \tablename\ \thetable{} -- continued}}
\\ \hline
\hline
Name & Type & $\log R'_{HK}$ & Observations & Ref.  \\
\hline \endhead

HD 142 & F7 & -5.020 & average of 2 individual points & 1  \\
       &    & -4.950 & individual on 2001 Aug 04 & 92 \\
& & & {\bf (P) Adopted value: -4.997 $\pm$ 0.060} \\
\hline
HD 1237 & G6 & -4.440 & 1992 individual & 5 \\
        &    & -4.270 & 61 obs in 2 years & 6 \\
        &    & -4.340 & average of above + extra individual & 1 \\
& & & {\bf (G) Adopted value: -4.273 $\pm$ 0.170} \\
\hline
HD 2039 & G2/3 &-4.980 & average of 2 individuals & 1 \\
\end{longtable}

\end{center}

\twocolumn

\bibliographystyle{mn2e}
\bibliography{abbrev,refs}

\begin{thebibliography}{}

\bibitem[\protect\citeauthoryear{{Acke} \& {Waelkens}}{{Acke} \&
  {Waelkens}}{2004}]{acke04}
{Acke} B.,  {Waelkens} C.,  2004, A\&A, 427, 1009

\bibitem[\protect\citeauthoryear{{Alonso}, {Brown}, {Torres}, {Latham},
  {Sozzetti}, {Mandushev}, {Belmonte}, {Charbonneau}, {Deeg}, {Dunham},
  {O'Donovan} \& {Stefanik}}{{Alonso} et~al.}{2004}]{alonso04}
{Alonso} R.,  {Brown} T.~M.,  {Torres} G.,  {Latham} D.~W.,  {Sozzetti} A.,
  {Mandushev} G.,  {Belmonte} J.~A.,  {Charbonneau} D.,  {Deeg} H.~J.,
  {Dunham} E.~W.,  {O'Donovan} F.~T.,    {Stefanik} R.~P.,  2004, ApJ, 613,
  L153

\bibitem[\protect\citeauthoryear{{Anderson}, {Hellier}, {Gillon}, {Triaud},
  {Smalley}, {Hebb}, {Collier Cameron}, {Maxted}, {Queloz}, {West}, {Bentley},
  {Enoch}, {Horne}, {Lister}, {Mayor}, {Parley}, {Pepe} \& {et al.}}{{Anderson}
  et~al.}{2010}]{anderson10}
{Anderson} D.~R.,  {Hellier} C.,  {Gillon} M.,  {Triaud} A. H. M.~J.,
  {Smalley} B.,  {Hebb} L.,  {Collier Cameron} A.,  {Maxted} P. F.~L.,
  {Queloz} D.,  {West} R.~G.,  {Bentley} S.~J.,  {Enoch} B.,  {Horne} K.,
  {Lister} T.~A.,  {Mayor} M.,  {Parley} N.~R.,  {Pepe} F.,    {et al.} 2010,
  ApJ, 709, 159

\bibitem[\protect\citeauthoryear{{Bakos}, {Kov{\'a}cs}, {Torres}, {Fischer},
  {Latham}, {Noyes}, {Sasselov}, {Mazeh}, {Shporer}, {Butler}, {Stefanik},
  {Fern{\'a}ndez}, {Sozzetti}, {P{\'a}l}, {Johnson}, {Marcy}, {Winn} \& {et
  al.}}{{Bakos} et~al.}{2007}]{bakos07b}
{Bakos} G.~{\'A}.,  {Kov{\'a}cs} G.,  {Torres} G.,  {Fischer} D.~A.,  {Latham}
  D.~W.,  {Noyes} R.~W.,  {Sasselov} D.~D.,  {Mazeh} T.,  {Shporer} A.,
  {Butler} R.~P.,  {Stefanik} R.~P.,  {Fern{\'a}ndez} J.~M.,  {Sozzetti} A.,
  {P{\'a}l} A.,  {Johnson} J.,  {Marcy} G.~W.,  {Winn} J.~N.,    {et al.} 2007,
  ApJ, 670, 826

\bibitem[\protect\citeauthoryear{Bakos, Noyes, Kov{\'a}cs, Latham, Sasselov,
  Torres, Fischer, Stefanik, Sato, Johnson, P{\'a}l, Marcy, Butler, Esquerdo,
  Stanek, L{\'a}z{\'a}r, Papp, S{\'a}ri \& Sip{\H o}cz}{Bakos
  et~al.}{2007}]{bakos07a}
Bakos G.~{\'A}.,  Noyes R.~W.,  Kov{\'a}cs G.,  Latham D.~W.,  Sasselov D.~D.,
  Torres G.,  Fischer D.~A.,  Stefanik R.~P.,  Sato B.,  Johnson J.~A.,
  P{\'a}l A.,  Marcy G.~W.,  Butler R.~P.,  Esquerdo G.~A.,  Stanek K.~Z.,
  L{\'a}z{\'a}r J.,  Papp I.,  S{\'a}ri P.,    Sip{\H o}cz B.,  2007, ApJ, 656,
  552

\bibitem[\protect\citeauthoryear{Barnes, Collier~Cameron, James \&
  Donati}{Barnes et~al.}{2001}]{barnes01}
Barnes J.~R.,  Collier~Cameron A.,  James D.,    Donati J.-F.,  2001, MNRAS,
  324, 231

\bibitem[\protect\citeauthoryear{Barnes}{Barnes}{2001}]{sbarnes01}
Barnes S.~A.,  2001, ApJ, 561, 1095

\bibitem[\protect\citeauthoryear{Barnes}{Barnes}{2007}]{barnes07}
Barnes S.~A.,  2007, ApJ, 669, 1167

\bibitem[\protect\citeauthoryear{Beck \& Giles}{Beck \& Giles}{2005}]{beck05}
Beck J.~G.,  Giles P.,  2005, ApJ, 621, L153

\bibitem[\protect\citeauthoryear{{Benz} \& {Mayor}}{{Benz} \&
  {Mayor}}{1984}]{benz84}
{Benz} W.,  {Mayor} M.,  1984, A\&A, 138, 183

\bibitem[\protect\citeauthoryear{{Bernacca} \& {Perinotto}}{{Bernacca} \&
  {Perinotto}}{1970}]{bernacca70}
{Bernacca} P.~L.,  {Perinotto} M.,  1970, Contributions dell'Osservatorio
  Astrofisica dell'Universita di Padova in Asiago, 239, 1

\bibitem[\protect\citeauthoryear{{Bernkopf}, {Fidler} \& {Fuhrmann}}{{Bernkopf}
  et~al.}{2001}]{bernkopf01}
{Bernkopf} J.,  {Fidler} A.,    {Fuhrmann} K.,  2001, in {von Hippel} T.,
  {Simpson} C.,   {Manset} N.,  eds, Astrophysical Ages and Times Scales
  Vol.~245 of Astronomical Society of the Pacific Conference Series, {The Dark
  Side of the Milky Way}.
pp 207--+

\bibitem[\protect\citeauthoryear{Beuermann, Baraffe \& Hauschildt}{Beuermann
  et~al.}{1999}]{beuermann99}
Beuermann K.,  Baraffe I.,    Hauschildt P.,  1999, A\&A, 348, 524

\bibitem[\protect\citeauthoryear{{Bouchy}, {Pont}, {Santos}, {Melo}, {Mayor},
  {Queloz} \& {Udry}}{{Bouchy} et~al.}{2004}]{bouchy04}
{Bouchy} F.,  {Pont} F.,  {Santos} N.~C.,  {Melo} C.,  {Mayor} M.,  {Queloz}
  D.,    {Udry} S.,  2004, A\&A, 421, L13

\bibitem[\protect\citeauthoryear{{Bouchy}, {Queloz}, {Deleuil}, {Loeillet},
  {Hatzes}, {Aigrain}, {Alonso}, {Auvergne}, {Baglin}, {Barge}, {Benz},
  {Bord{\'e}}, {Deeg}, {de La Reza}, {Dvorak}, {Erikson}, {Fridlund}, {Gondoin}
  \& {et al.}}{{Bouchy} et~al.}{2008}]{bouchy08}
{Bouchy} F.,  {Queloz} D.,  {Deleuil} M.,  {Loeillet} B.,  {Hatzes} A.~P.,
  {Aigrain} S.,  {Alonso} R.,  {Auvergne} M.,  {Baglin} A.,  {Barge} P.,
  {Benz} W.,  {Bord{\'e}} P.,  {Deeg} H.~J.,  {de La Reza} R.,  {Dvorak} R.,
  {Erikson} A.,  {Fridlund} M.,  {Gondoin} P.,    {et al.} 2008, A\&A, 482, L25

\bibitem[\protect\citeauthoryear{{Bouchy}, {Udry}, {Mayor}, {Moutou}, {Pont},
  {Iribarne}, {da Silva}, {Ilovaisky}, {Queloz}, {Santos}, {S{\'e}gransan} \&
  {Zucker}}{{Bouchy} et~al.}{2005}]{bouchy05}
{Bouchy} F.,  {Udry} S.,  {Mayor} M.,  {Moutou} C.,  {Pont} F.,  {Iribarne} N.,
   {da Silva} R.,  {Ilovaisky} S.,  {Queloz} D.,  {Santos} N.~C.,
  {S{\'e}gransan} D.,    {Zucker} S.,  2005, A\&A, 444, L15

\bibitem[\protect\citeauthoryear{Brown}{Brown}{2010}]{brown10}
Brown T.~M.,  2010, ApJ, 709, 535

\bibitem[\protect\citeauthoryear{Butler, Marcy, Vogt, Fischer, Henry, Laughlin
  \& Wright}{Butler et~al.}{2003}]{butler03}
Butler R.~P.,  Marcy G.~W.,  Vogt S.~S.,  Fischer D.~A.,  Henry G.~W.,
  Laughlin G.,    Wright J.~T.,  2003, ApJ, 582, 455

\bibitem[\protect\citeauthoryear{Butler, Vogt, Marcy, Fischer, Henry \&
  Apps}{Butler et~al.}{2000}]{butler00}
Butler R.~P.,  Vogt S.~S.,  Marcy G.~W.,  Fischer D.~A.,  Henry G.~W.,    Apps
  K.,  2000, ApJ, 545, 504

\bibitem[\protect\citeauthoryear{Butler, Wright, Marcy, Fischer, Vogt, Tinney,
  Jones, Carter, Johnson, McCarthy \& Penny}{Butler et~al.}{2006}]{butler06}
Butler R.~P.,  Wright J.~T.,  Marcy G.~W.,  Fischer D.~A.,  Vogt S.~S.,  Tinney
  C.~G.,  Jones H. R.~A.,  Carter B.~D.,  Johnson J.~A.,  McCarthy C.,    Penny
  A.~J.,  2006, ApJ, 646, 505

\bibitem[\protect\citeauthoryear{Cameron \& Donati}{Cameron \&
  Donati}{2002}]{cameron02}
Cameron A.~C.,  Donati J.-F.,  2002, MNRAS, 329, 23

\bibitem[\protect\citeauthoryear{Cameron \& Foing}{Cameron \&
  Foing}{1997}]{cameron97}
Cameron A.~C.,  Foing B.~H.,  1997, Observatory, 117, 218

\bibitem[\protect\citeauthoryear{Cochran, Redfield, Endl \& Cochran}{Cochran
  et~al.}{2008}]{cochran08}
Cochran W.~D.,  Redfield S.,  Endl M.,    Cochran A.~L.,  2008, ApJ, 683, L59

\bibitem[\protect\citeauthoryear{{Collier Cameron}, {Wilson}, {West}, {Hebb},
  {Wang}, {Aigrain}, {Bouchy}, {Christian}, {Clarkson}, {Enoch}, {Esposito},
  {Guenther}, {Haswell}, {H{\'e}brard}, {Hellier}, {Horne} \& {et
  al.}}{{Collier Cameron} et~al.}{2007}]{cameron07}
{Collier Cameron} A.,  {Wilson} D.~M.,  {West} R.~G.,  {Hebb} L.,  {Wang}
  X.-B.,  {Aigrain} S.,  {Bouchy} F.,  {Christian} D.~J.,  {Clarkson} W.~I.,
  {Enoch} B.,  {Esposito} M.,  {Guenther} E.,  {Haswell} C.~A.,  {H{\'e}brard}
  G.,  {Hellier} C.,  {Horne} K.,    {et al.} 2007, MNRAS, 380, 1230

\bibitem[\protect\citeauthoryear{Da~Silva, Udry, Bouchy, Mayor, Moutou, Pont,
  Queloz, Santos, S\'{e}gransan \& Zucker}{Da~Silva et~al.}{2006}]{dasilva06}
Da~Silva R.,  Udry S.,  Bouchy F.,  Mayor M.,  Moutou C.,  Pont F.,  Queloz D.,
   Santos N.~C.,  S\'{e}gransan D.,    Zucker S.,  2006, A\&A, 446, 717

\bibitem[\protect\citeauthoryear{{de Medeiros} \& {Mayor}}{{de Medeiros} \&
  {Mayor}}{1999}]{demedeiros99}
{de Medeiros} J.~R.,  {Mayor} M.,  1999, A\&AS, 139, 433

\bibitem[\protect\citeauthoryear{{di Benedetto} \& {Rabbia}}{{di Benedetto} \&
  {Rabbia}}{1987}]{diBenedetto87}
{di Benedetto} G.~P.,  {Rabbia} Y.,  1987, A\&A, 188, 114

\bibitem[\protect\citeauthoryear{Eggenberger, Mayor, Naef, Pepe, Queloz,
  Santos, Udry \& Lovis}{Eggenberger et~al.}{2006}]{eggenberger06}
Eggenberger A.,  Mayor M.,  Naef D.,  Pepe F.,  Queloz D.,  Santos N.~C.,  Udry
  S.,    Lovis C.,  2006, A\&A, 447, 1159

\bibitem[\protect\citeauthoryear{{Fischer}, {Laughlin}, {Marcy}, {Butler},
  {Vogt}, {Johnson}, {Henry}, {McCarthy}, {Ammons}, {Robinson}, {Strader},
  {Valenti}, {McCullough}, {Charbonneau}, {Haislip}, {Knutson}, {Reichart} \&
  {et. al}}{{Fischer} et~al.}{2006}]{fischer06}
{Fischer} D.~A.,  {Laughlin} G.,  {Marcy} G.~W.,  {Butler} R.~P.,  {Vogt}
  S.~S.,  {Johnson} J.~A.,  {Henry} G.~W.,  {McCarthy} C.,  {Ammons} M.,
  {Robinson} S.,  {Strader} J.,  {Valenti} J.~A.,  {McCullough} P.~R.,
  {Charbonneau} D.,  {Haislip} J.,  {Knutson} H.~A.,  {Reichart} D.~E.,    {et.
  al} 2006, ApJ, 637, 1094

\bibitem[\protect\citeauthoryear{{Fischer}, {Marcy}, {Butler}, {Vogt}, {Frink}
  \& {Apps}}{{Fischer} et~al.}{2001}]{fischer01}
{Fischer} D.~A.,  {Marcy} G.~W.,  {Butler} R.~P.,  {Vogt} S.~S.,  {Frink} S.,
   {Apps} K.,  2001, ApJ, 551, 1107

\bibitem[\protect\citeauthoryear{{Fischer}, {Marcy}, {Butler}, {Vogt}, {Walp}
  \& {Apps}}{{Fischer} et~al.}{2002}]{fischer02}
{Fischer} D.~A.,  {Marcy} G.~W.,  {Butler} R.~P.,  {Vogt} S.~S.,  {Walp} B.,
  {Apps} K.,  2002, PASP, 114, 529

\bibitem[\protect\citeauthoryear{Fischer \& Valenti}{Fischer \&
  Valenti}{2005}]{fischer05}
Fischer D.~A.,  Valenti J.,  2005, ApJ, 622, 1102

\bibitem[\protect\citeauthoryear{{Fischer}, {Vogt}, {Marcy}, {Butler}, {Sato},
  {Henry}, {Robinson}, {Laughlin}, {Ida}, {Toyota}, {Omiya}, {Driscoll},
  {Takeda}, {Wright} \& {Johnson}}{{Fischer} et~al.}{2007}]{fischer07}
{Fischer} D.~A.,  {Vogt} S.~S.,  {Marcy} G.~W.,  {Butler} R.~P.,  {Sato} B.,
  {Henry} G.~W.,  {Robinson} S.,  {Laughlin} G.,  {Ida} S.,  {Toyota} E.,
  {Omiya} M.,  {Driscoll} P.,  {Takeda} G.,  {Wright} J.~T.,    {Johnson}
  J.~A.,  2007, ApJ, 669, 1336

\bibitem[\protect\citeauthoryear{Ford}{Ford}{2006}]{ford06}
Ford E.~B.,  2006, ApJ, 642, 505

\bibitem[\protect\citeauthoryear{{Fouque} \& {Gieren}}{{Fouque} \&
  {Gieren}}{1997}]{fouque97}
{Fouque} P.,  {Gieren} W.~P.,  1997, A\&A, 320, 799

\bibitem[\protect\citeauthoryear{Fracassini, Pastori, Covino \&
  Pozzi}{Fracassini et~al.}{2001}]{fracassini01}
Fracassini L. E.~P.,  Pastori L.,  Covino S.,    Pozzi A.,  2001, A\&A, 367,
  521

\bibitem[\protect\citeauthoryear{Fuhrmann}{Fuhrmann}{1998}]{fuhrmann98b}
Fuhrmann K.,  1998, A\&A, 338, 161

\bibitem[\protect\citeauthoryear{Fuhrmann, Pfeiffer \& Bernkopf}{Fuhrmann
  et~al.}{1997}]{fuhrmann97}
Fuhrmann K.,  Pfeiffer M.~J.,    Bernkopf J.,  1997, A\&A, 326, 1081

\bibitem[\protect\citeauthoryear{Fuhrmann, Pfeiffer \& Bernkopf}{Fuhrmann
  et~al.}{1998}]{fuhrmann98}
Fuhrmann K.,  Pfeiffer M.~J.,    Bernkopf J.,  1998, A\&A, 336, 942

\bibitem[\protect\citeauthoryear{{Galland}, {Lagrange}, {Udry}, {Chelli},
  {Pepe}, {Beuzit} \& {Mayor}}{{Galland} et~al.}{2005}]{galland05}
{Galland} F.,  {Lagrange} A.-M.,  {Udry} S.,  {Chelli} A.,  {Pepe} F.,
  {Beuzit} J.-L.,    {Mayor} M.,  2005, A\&A, 444, L21

\bibitem[\protect\citeauthoryear{Gaudi \& Winn}{Gaudi \& Winn}{2007}]{gaudi07}
Gaudi B.~S.,  Winn J.~N.,  2007, ApJ, 655, 550

\bibitem[\protect\citeauthoryear{{Ge}, {van Eyken}, {Mahadevan}, {DeWitt},
  {Kane}, {Cohen}, {Vanden Heuvel}, {Fleming}, {Guo}, {Henry}, {Schneider},
  {Ramsey}, {Wittenmyer}, {Endl}, {Cochran} \& {et al.}}{{Ge}
  et~al.}{2006}]{ge06}
{Ge} J.,  {van Eyken} J.,  {Mahadevan} S.,  {DeWitt} C.,  {Kane} S.~R.,
  {Cohen} R.,  {Vanden Heuvel} A.,  {Fleming} S.~W.,  {Guo} P.,  {Henry} G.~W.,
   {Schneider} D.~P.,  {Ramsey} L.~W.,  {Wittenmyer} R.~A.,  {Endl} M.,
  {Cochran} W.~D.,    {et al.} 2006, ApJ, 648, 683

\bibitem[\protect\citeauthoryear{{Gillon}, {Anderson}, {Triaud}, {Hellier},
  {Maxted}, {Pollaco}, {Queloz}, {Smalley}, {West}, {Wilson}, {Bentley},
  {Collier Cameron}, {Enoch}, {Hebb}, {Horne}, {Irwin}, {Joshi}, {Lister} \& et
  al.}{{Gillon} et~al.}{2009}]{gillon09}
{Gillon} M.,  {Anderson} D.~R.,  {Triaud} A.~H.~M.~J.,  {Hellier} C.,  {Maxted}
  P.~F.~L.,  {Pollaco} D.,  {Queloz} D.,  {Smalley} B.,  {West} R.~G.,
  {Wilson} D.~M.,  {Bentley} S.~J.,  {Collier Cameron} A.,  {Enoch} B.,  {Hebb}
  L.,  {Horne} K.,  {Irwin} J.,  {Joshi} Y.~C.,  {Lister} T.~A.,    et al.
  2009, A\&A, 501, 785

\bibitem[\protect\citeauthoryear{Gizon \& Solanki}{Gizon \&
  Solanki}{2003}]{gizon03}
Gizon L.,  Solanki S.~K.,  2003, ApJ, 589, 1009

\bibitem[\protect\citeauthoryear{Gonzalez}{Gonzalez}{1998}]{gonzalez98}
Gonzalez G.,  1998, A\&A, 334, 221

\bibitem[\protect\citeauthoryear{Gregory}{Gregory}{2007}]{gregory07}
Gregory P.~C.,  2007, MNRAS, 374, 1321

\bibitem[\protect\citeauthoryear{Hartigan \& Hartigan}{Hartigan \&
  Hartigan}{1985}]{hartigan85}
Hartigan J.~A.,  Hartigan P.~M.,  1985, Annals of Statistics, 13, 70

\bibitem[\protect\citeauthoryear{{Hatzes}, {Cochran}, {Endl}, {Guenther},
  {Saar}, {Walker}, {Yang}, {Hartmann}, {Esposito}, {Paulson} \&
  {D{\"o}llinger}}{{Hatzes} et~al.}{2006}]{hatzes06}
{Hatzes} A.~P.,  {Cochran} W.~D.,  {Endl} M.,  {Guenther} E.~W.,  {Saar} S.~H.,
   {Walker} G.~A.~H.,  {Yang} S.,  {Hartmann} M.,  {Esposito} M.,  {Paulson}
  D.~B.,    {D{\"o}llinger} M.~P.,  2006, A\&A, 457, 335

\bibitem[\protect\citeauthoryear{H\'{e}brard, Bouchy, Pont, Loeillet, Rabus,
  Bonfils, Moutou, Boisse, Delfosse, Desort, Eggenberger, Ehrenreich,
  Forveille, Lagrange, Lovis, Mayor, Pepe, Perrier, Queloz, Santos \& et
  al.}{H\'{e}brard et~al.}{2008}]{hebrard08}
H\'{e}brard G.,  Bouchy F.,  Pont F.,  Loeillet B.,  Rabus M.,  Bonfils X.,
  Moutou C.,  Boisse I.,  Delfosse X.,  Desort M.,  Eggenberger A.,  Ehrenreich
  D.,  Forveille T.,  Lagrange A.-M.,  Lovis C.,  Mayor M.,  Pepe F.,  Perrier
  C.,  Queloz D.,  Santos N.~C.,    et al. 2008, A\&A, 488, 763

\bibitem[\protect\citeauthoryear{Henry, Baliunas, Donahue, Fekel \& Soon}{Henry
  et~al.}{2002}]{henry02b}
Henry G.~W.,  Baliunas S.~L.,  Donahue R.~A.,  Fekel F.~C.,    Soon W.,  2002,
  AJ, 531, 415

\bibitem[\protect\citeauthoryear{{Henry}, {Donahue} \& {Baliunas}}{{Henry}
  et~al.}{2002}]{henry02}
{Henry} G.~W.,  {Donahue} R.~A.,    {Baliunas} S.~L.,  2002, ApJ, 577, L111

\bibitem[\protect\citeauthoryear{Henry, Soderblom, Donahue \& Baliunas}{Henry
  et~al.}{1996}]{henry96}
Henry T.~J.,  Soderblom D.~R.,  Donahue R.~A.,    Baliunas S.~L.,  1996, AJ,
  111, 439

\bibitem[\protect\citeauthoryear{{Irwin}, {Charbonneau}, {Nutzman}, {Welsh},
  {Rajan}, {Hidas}, {Brown}, {Lister}, {Davies}, {Laughlin} \&
  {Langton}}{{Irwin} et~al.}{2008}]{irwin08}
{Irwin} J.,  {Charbonneau} D.,  {Nutzman} P.,  {Welsh} W.~F.,  {Rajan} A.,
  {Hidas} M.,  {Brown} T.~M.,  {Lister} T.~A.,  {Davies} D.,  {Laughlin} G.,
  {Langton} J.,  2008, ApJ, 681, 636

\bibitem[\protect\citeauthoryear{{Jenkins}, {Borucki}, {Koch}, {Marcy},
  {Cochran}, {Basri}, {Batalha}, {Buchhave}, {Brown}, {Caldwell}, {Dunham},
  {Endl}, {Fischer}, {Gautier} III, {Geary}, {Gilliland}, {Howell}, {Isaacson}
  \& et al.}{{Jenkins} et~al.}{2010}]{jenkins10}
{Jenkins} J.~M.,  {Borucki} W.~J.,  {Koch} D.~G.,  {Marcy} G.~W.,  {Cochran}
  W.~D.,  {Basri} G.,  {Batalha} N.~M.,  {Buchhave} L.~A.,  {Brown} T.~M.,
  {Caldwell} D.~A.,  {Dunham} E.~W.,  {Endl} M.,  {Fischer} D.~A.,  {Gautier}
  III T.~N.,  {Geary} J.~C.,  {Gilliland} R.~L.,  {Howell} S.~B.,  {Isaacson}
  H.,    et al. 2010, arXiv1001.0416

\bibitem[\protect\citeauthoryear{Jenkins, Jones, Tinney, Butler, McCarthy,
  Marcy, Pinfield, Carter \& Penny}{Jenkins et~al.}{2006}]{jenkins06}
Jenkins J.~S.,  Jones H. R.~A.,  Tinney C.~G.,  Butler R.~P.,  McCarthy C.,
  Marcy G.~W.,  Pinfield D.~J.,  Carter B.~D.,    Penny A.~J.,  2006, MNRAS,
  3372, 163

\bibitem[\protect\citeauthoryear{{Johnson}, {Fischer}, {Marcy}, {Wright},
  {Driscoll}, {Butler}, {Hekker}, {Reffert} \& {Vogt}}{{Johnson}
  et~al.}{2007}]{johnson07}
{Johnson} J.~A.,  {Fischer} D.~A.,  {Marcy} G.~W.,  {Wright} J.~T.,  {Driscoll}
  P.,  {Butler} R.~P.,  {Hekker} S.,  {Reffert} S.,    {Vogt} S.~S.,  2007,
  ApJ, 665, 785

\bibitem[\protect\citeauthoryear{Johnson, Marcy, Fischer, Henry, Wright,
  Isaacson \& McCarthy}{Johnson et~al.}{2006}]{johnson06b}
Johnson J.~A.,  Marcy G.~W.,  Fischer D.~A.,  Henry G.~W.,  Wright J.~T.,
  Isaacson H.,    McCarthy C.,  2006, ApJ, 652, 1724

\bibitem[\protect\citeauthoryear{Johnson, Marcy, Fischer, Laughlin, Butler,
  Henry, Valenti, Ford, Vogt \& Wright}{Johnson et~al.}{2006}]{johnson06a}
Johnson J.~A.,  Marcy G.~W.,  Fischer D.~A.,  Laughlin G.,  Butler R.~P.,
  Henry G.~W.,  Valenti J.~A.,  Ford E.~B.,  Vogt S.~S.,    Wright J.~T.,
  2006, ApJ, 647, 600

\bibitem[\protect\citeauthoryear{Johnson, Winn, Albrecht, Howard, Marcy \&
  Gazak}{Johnson et~al.}{2009}]{johnson09}
Johnson J.~A.,  Winn J.~N.,  Albrecht S.,  Howard A.~W.,  Marcy G.~W.,    Gazak
  J.~Z.,  2009, PASP, 121, 1104

\bibitem[\protect\citeauthoryear{Johnson, Winn, Narita, Enya, Williams, Marcy,
  Sato, Ohta, Taruya, Suto, Turner, Bakos, Butler, Vogt, Aoki, Tamura, Yamada,
  Yoshii \& Hidas}{Johnson et~al.}{2008}]{johnson08}
Johnson J.~A.,  Winn J.~N.,  Narita N.,  Enya K.,  Williams P. K.~G.,  Marcy
  G.~W.,  Sato B.,  Ohta Y.,  Taruya A.,  Suto Y.,  Turner E.~L.,  Bakos G.,
  Butler R.~P.,  Vogt S.~S.,  Aoki W.,  Tamura M.,  Yamada T.,  Yoshii Y.,
  Hidas M.,  2008, ApJ, 686, 649

\bibitem[\protect\citeauthoryear{Jones, Butler, Tinney, Marcy, Penny, McCarthy
  \& Carter}{Jones et~al.}{2003}]{jones03}
Jones H. R.~A.,  Butler P.~R.,  Tinney C.~G.,  Marcy G.~W.,  Penny A.~J.,
  McCarthy C.,    Carter B.~D.,  2003, MNRAS, 341, 948

\bibitem[\protect\citeauthoryear{{Jones}, {Butler}, {Tinney}, {Marcy},
  {Carter}, {Penny}, {McCarthy} \& {Bailey}}{{Jones} et~al.}{2006}]{jones06}
{Jones} H.~R.~A.,  {Butler} R.~P.,  {Tinney} C.~G.,  {Marcy} G.~W.,  {Carter}
  B.~D.,  {Penny} A.~J.,  {McCarthy} C.,    {Bailey} J.,  2006, MNRAS, 369, 249

\bibitem[\protect\citeauthoryear{Jorissen, Mayor \& Udry}{Jorissen
  et~al.}{2001}]{jorissen01}
Jorissen A.,  Mayor M.,    Udry S.,  2001, A\&A, 379, 992

\bibitem[\protect\citeauthoryear{{Konacki}, {Torres}, {Sasselov} \&
  {Jha}}{{Konacki} et~al.}{2005}]{konacki05}
{Konacki} M.,  {Torres} G.,  {Sasselov} D.~D.,    {Jha} S.,  2005, ApJ, 624,
  372

\bibitem[\protect\citeauthoryear{{Konacki}, {Torres}, {Sasselov},
  {Pietrzy{\'n}ski}, {Udalski}, {Jha}, {Ruiz}, {Gieren} \& {Minniti}}{{Konacki}
  et~al.}{2004}]{konacki04}
{Konacki} M.,  {Torres} G.,  {Sasselov} D.~D.,  {Pietrzy{\'n}ski} G.,
  {Udalski} A.,  {Jha} S.,  {Ruiz} M.~T.,  {Gieren} W.,    {Minniti} D.,  2004,
  ApJ, 609, L37

\bibitem[\protect\citeauthoryear{{Korzennik}, {Brown}, {Fischer}, {Nisenson} \&
  {Noyes}}{{Korzennik} et~al.}{2000}]{korzennik00}
{Korzennik} S.~G.,  {Brown} T.~M.,  {Fischer} D.~A.,  {Nisenson} P.,    {Noyes}
  R.~W.,  2000, ApJ, 533, L147

\bibitem[\protect\citeauthoryear{{Laughlin}, {Wolf}, {Vanmunster},
  {Bodenheimer}, {Fischer}, {Marcy}, {Butler} \& {Vogt}}{{Laughlin}
  et~al.}{2005}]{laughlin05}
{Laughlin} G.,  {Wolf} A.,  {Vanmunster} T.,  {Bodenheimer} P.,  {Fischer} D.,
  {Marcy} G.,  {Butler} P.,    {Vogt} S.,  2005, ApJ, 621, 1072

\bibitem[\protect\citeauthoryear{{Lo Curto}, {Mayor}, {Clausen}, {Benz},
  {Bouchy}, {Lovis}, {Moutou}, {Naef}, {Pepe}, {Queloz}, {Santos}, {Sivan},
  {Udry}, {Bonfils}, {Delfosse}, {Mordasini}, {Fouqu{\'e}}, {Olsen} \&
  {Pritchard}}{{Lo Curto} et~al.}{2006}]{locurto06}
{Lo Curto} G.,  {Mayor} M.,  {Clausen} J.~V.,  {Benz} W.,  {Bouchy} F.,
  {Lovis} C.,  {Moutou} C.,  {Naef} D.,  {Pepe} F.,  {Queloz} D.,  {Santos}
  N.~C.,  {Sivan} J.-P.,  {Udry} S.,  {Bonfils} X.,  {Delfosse} X.,
  {Mordasini} C.,  {Fouqu{\'e}} P.,  {Olsen} E.~H.,    {Pritchard} J.~D.,
  2006, A\&A, 451, 345

\bibitem[\protect\citeauthoryear{{Lovis}, {Mayor}, {Bouchy}, {Pepe}, {Queloz},
  {Santos}, {Udry}, {Benz}, {Bertaux}, {Mordasini} \& {Sivan}}{{Lovis}
  et~al.}{2005}]{lovis05}
{Lovis} C.,  {Mayor} M.,  {Bouchy} F.,  {Pepe} F.,  {Queloz} D.,  {Santos}
  N.~C.,  {Udry} S.,  {Benz} W.,  {Bertaux} J.-L.,  {Mordasini} C.,    {Sivan}
  J.-P.,  2005, A\&A, 437, 1121

\bibitem[\protect\citeauthoryear{{Lovis}, {Mayor}, {Pepe}, {Alibert}, {Benz},
  {Bouchy}, {Correia}, {Laskar}, {Mordasini}, {Queloz}, {Santos}, {Udry},
  {Bertaux} \& {Sivan}}{{Lovis} et~al.}{2006}]{lovis06}
{Lovis} C.,  {Mayor} M.,  {Pepe} F.,  {Alibert} Y.,  {Benz} W.,  {Bouchy} F.,
  {Correia} A.~C.~M.,  {Laskar} J.,  {Mordasini} C.,  {Queloz} D.,  {Santos}
  N.~C.,  {Udry} S.,  {Bertaux} J.-L.,    {Sivan} J.-P.,  2006, Nature, 441,
  305

\bibitem[\protect\citeauthoryear{{Lowrance}, {Becklin}, {Schneider},
  {Kirkpatrick}, {Weinberger}, {Zuckerman}, {Dumas}, {Beuzit}, {Plait},
  {Malumuth}, {Heap}, {Terrile} \& {Hines}}{{Lowrance}
  et~al.}{2005}]{lowrance05}
{Lowrance} P.~J.,  {Becklin} E.~E.,  {Schneider} G.,  {Kirkpatrick} J.~D.,
  {Weinberger} A.~J.,  {Zuckerman} B.,  {Dumas} C.,  {Beuzit} J.-L.,  {Plait}
  P.,  {Malumuth} E.,  {Heap} S.,  {Terrile} R.~J.,    {Hines} D.~C.,  2005,
  AJ, 130, 1845

\bibitem[\protect\citeauthoryear{Malmberg, Davies \& Chambers}{Malmberg
  et~al.}{2007}]{malmberg07}
Malmberg D.,  Davies M.~B.,    Chambers J.~E.,  2007, MNRAS, 377, L1

\bibitem[\protect\citeauthoryear{{Marcy}, {Butler}, {Fischer}, {Laughlin},
  {Vogt}, {Henry} \& {Pourbaix}}{{Marcy} et~al.}{2002}]{marcy02}
{Marcy} G.~W.,  {Butler} R.~P.,  {Fischer} D.~A.,  {Laughlin} G.,  {Vogt}
  S.~S.,  {Henry} G.~W.,    {Pourbaix} D.,  2002, ApJ, 581, 1375

\bibitem[\protect\citeauthoryear{{Masana}, {Jordi} \& {Ribas}}{{Masana}
  et~al.}{2006}]{masana06}
{Masana} E.,  {Jordi} C.,    {Ribas} I.,  2006, A\&A, 450, 735

\bibitem[\protect\citeauthoryear{Mayor \& Queloz}{Mayor \&
  Queloz}{1995}]{mayor95}
Mayor M.,  Queloz D.,  1995, Nature, 378, 355

\bibitem[\protect\citeauthoryear{Mayor, Udry, Naef, Pepe, Queloz, Santos \&
  Burnet}{Mayor et~al.}{2004}]{mayor04}
Mayor M.,  Udry S.,  Naef D.,  Pepe F.,  Queloz D.,  Santos N.~C.,    Burnet
  M.,  2004, A\&A, 415, 391

\bibitem[\protect\citeauthoryear{{Mazeh}, {Naef}, {Torres}, {Latham}, {Mayor},
  {Beuzit}, {Brown}, {Buchhave}, {Burnet}, {Carney}, {Charbonneau}, {Drukier},
  {Laird}, {Pepe}, {Perrier}, {Queloz}, {Santos}, {Sivan}, {Udry} \&
  {Zucker}}{{Mazeh} et~al.}{2000}]{mazeh00}
{Mazeh} T.,  {Naef} D.,  {Torres} G.,  {Latham} D.~W.,  {Mayor} M.,  {Beuzit}
  J.-L.,  {Brown} T.~M.,  {Buchhave} L.,  {Burnet} M.,  {Carney} B.~W.,
  {Charbonneau} D.,  {Drukier} G.~A.,  {Laird} J.~B.,  {Pepe} F.,  {Perrier}
  C.,  {Queloz} D.,  {Santos} N.~C.,  {Sivan} J.-P.,  {Udry} S.,    {Zucker}
  S.,  2000, ApJ, 532, L55

\bibitem[\protect\citeauthoryear{{Melo}, {Santos}, {Gieren}, {Pietrzynski},
  {Ruiz}, {Sousa}, {Bouchy}, {Lovis}, {Mayor}, {Pepe}, {Queloz}, {da Silva} \&
  {Udry}}{{Melo} et~al.}{2007}]{melo07}
{Melo} C.,  {Santos} N.~C.,  {Gieren} W.,  {Pietrzynski} G.,  {Ruiz} M.~T.,
  {Sousa} S.~G.,  {Bouchy} F.,  {Lovis} C.,  {Mayor} M.,  {Pepe} F.,  {Queloz}
  D.,  {da Silva} R.,    {Udry} S.,  2007, A\&A, 467, 721

\bibitem[\protect\citeauthoryear{{Melo}, {Santos}, {Pont}, {Guillot},
  {Israelian}, {Mayor}, {Queloz} \& {Udry}}{{Melo} et~al.}{2006}]{melo06}
{Melo} C.,  {Santos} N.~C.,  {Pont} F.,  {Guillot} T.,  {Israelian} G.,
  {Mayor} M.,  {Queloz} D.,    {Udry} S.,  2006, A\&A, 460, 251

\bibitem[\protect\citeauthoryear{{Messina}, {Rodon{\`o}} \& {Guinan}}{{Messina}
  et~al.}{2001}]{messina01}
{Messina} S.,  {Rodon{\`o}} M.,    {Guinan} E.~F.,  2001, A\&A, 366, 215

\bibitem[\protect\citeauthoryear{Moutou, Mayor, Bouchy, Lovis, Pepe, Queloz,
  Santos, Udry, Benz, Lo~Curto, Naef, S\'{e}gransan \& Sivan}{Moutou
  et~al.}{2005}]{moutou05}
Moutou C.,  Mayor M.,  Bouchy F.,  Lovis C.,  Pepe F.,  Queloz D.,  Santos
  N.~C.,  Udry S.,  Benz W.,  Lo~Curto G.,  Naef D.,  S\'{e}gransan D.,
  Sivan J.-P.,  2005, A\&A, 439, 367

\bibitem[\protect\citeauthoryear{{Naef}, {Latham}, {Mayor}, {Mazeh}, {Beuzit},
  {Drukier}, {Perrier-Bellet}, {Queloz}, {Sivan}, {Torres}, {Udry} \&
  {Zucker}}{{Naef} et~al.}{2001}]{naef01}
{Naef} D.,  {Latham} D.~W.,  {Mayor} M.,  {Mazeh} T.,  {Beuzit} J.~L.,
  {Drukier} G.~A.,  {Perrier-Bellet} C.,  {Queloz} D.,  {Sivan} J.~P.,
  {Torres} G.,  {Udry} S.,    {Zucker} S.,  2001, A\&A, 375, L27

\bibitem[\protect\citeauthoryear{{Naef}, {Mayor}, {Benz}, {Bouchy}, {Lo Curto},
  {Lovis}, {Moutou}, {Pepe}, {Queloz}, {Santos} \& {Udry}}{{Naef}
  et~al.}{2007}]{naef07}
{Naef} D.,  {Mayor} M.,  {Benz} W.,  {Bouchy} F.,  {Lo Curto} G.,  {Lovis} C.,
  {Moutou} C.,  {Pepe} F.,  {Queloz} D.,  {Santos} N.~C.,    {Udry} S.,  2007,
  A\&A, 470, 721

\bibitem[\protect\citeauthoryear{Naef, Mayor, Beuzit, Perrier, Queloz, Sivan \&
  Udry}{Naef et~al.}{2004}]{naef04}
Naef D.,  Mayor M.,  Beuzit J.~L.,  Perrier C.,  Queloz D.,  Sivan J.~P.,
  Udry S.,  2004, A\&A, 414, 351

\bibitem[\protect\citeauthoryear{{Naef}, {Mayor}, {Korzennik}, {Queloz},
  {Udry}, {Nisenson}, {Noyes}, {Brown}, {Beuzit}, {Perrier} \& {Sivan}}{{Naef}
  et~al.}{2003}]{naef03}
{Naef} D.,  {Mayor} M.,  {Korzennik} S.~G.,  {Queloz} D.,  {Udry} S.,
  {Nisenson} P.,  {Noyes} R.~W.,  {Brown} T.~M.,  {Beuzit} J.~L.,  {Perrier}
  C.,    {Sivan} J.~P.,  2003, A\&A, 410, 1051

\bibitem[\protect\citeauthoryear{{Narita}, {Enya}, {Sato}, {Ohta}, {Winn},
  {Suto}, {Taruya}, {Turner}, {Aoki}, {Yoshii}, {Yamada} \& {Tamura}}{{Narita}
  et~al.}{2007}]{narita07}
{Narita} N.,  {Enya} K.,  {Sato} B.,  {Ohta} Y.,  {Winn} J.~N.,  {Suto} Y.,
  {Taruya} A.,  {Turner} E.~L.,  {Aoki} W.,  {Yoshii} M.,  {Yamada} T.,
  {Tamura} Y.,  2007, PASJ, 59, 763

\bibitem[\protect\citeauthoryear{{Narita}, {Sato}, {Hirano} \&
  {Tamura}}{{Narita} et~al.}{2009}]{narita09}
{Narita} N.,  {Sato} B.,  {Hirano} T.,    {Tamura} M.,  2009, PASJ, 61, L35

\bibitem[\protect\citeauthoryear{{Narita}, {Sato}, {Hirano}, {Winn}, {Aoki} \&
  {Tamura}}{{Narita} et~al.}{2010}]{narita10}
{Narita} N.,  {Sato} B.,  {Hirano} T.,  {Winn} J.~N.,  {Aoki} W.,    {Tamura}
  M.,  2010, arXiv1003.2268

\bibitem[\protect\citeauthoryear{{Narita}, {Sato}, Oshima \& {Winn}}{{Narita}
  et~al.}{2008}]{narita08}
{Narita} N.,  {Sato} B.,  Oshima O.,    {Winn} J.~N.,  2008, PASJ, 60, 1

\bibitem[\protect\citeauthoryear{Nordstr{\"{o}}m, Mayor, Andersen, Holmberg,
  Pont, Jorgensen, Olsen, Udry \& Mowlavi}{Nordstr{\"{o}}m
  et~al.}{2004}]{nordstroem04}
Nordstr{\"{o}}m B.,  Mayor M.,  Andersen J.,  Holmberg J.,  Pont F.,  Jorgensen
  B.~R.,  Olsen E.~H.,  Udry S.,    Mowlavi N.,  2004, A\&A, 418, 989

\bibitem[\protect\citeauthoryear{Noyes, Hartmann, Baliunas, Duncan \&
  Vaughan}{Noyes et~al.}{1984}]{noyes84}
Noyes R.~W.,  Hartmann L.~W.,  Baliunas S.~L.,  Duncan D.~K.,    Vaughan A.~H.,
   1984, ApJ, 279, 763

\bibitem[\protect\citeauthoryear{{O'Toole}, {Butler}, {Tinney}, {Jones},
  {Marcy}, {Carter}, {McCarthy}, {Bailey}, {Penny}, {Apps} \&
  {Fischer}}{{O'Toole} et~al.}{2007}]{otoole07}
{O'Toole} S.~J.,  {Butler} R.~P.,  {Tinney} C.~G.,  {Jones} H.~R.~A.,  {Marcy}
  G.~W.,  {Carter} B.,  {McCarthy} C.,  {Bailey} J.,  {Penny} A.~J.,  {Apps}
  K.,    {Fischer} D.,  2007, ApJ, 660, 1636

\bibitem[\protect\citeauthoryear{{Pepe}, {Mayor}, {Galland}, {Naef}, {Queloz},
  {Santos}, {Udry} \& {Burnet}}{{Pepe} et~al.}{2002}]{pepe02}
{Pepe} F.,  {Mayor} M.,  {Galland} F.,  {Naef} D.,  {Queloz} D.,  {Santos}
  N.~C.,  {Udry} S.,    {Burnet} M.,  2002, A\&A, 388, 632

\bibitem[\protect\citeauthoryear{{Pepe}, {Mayor}, {Queloz}, {Benz}, {Bonfils},
  {Bouchy}, {Curto}, {Lovis}, {M{\'e}gevand}, {Moutou}, {Naef}, {Rupprecht},
  {Santos}, {Sivan}, {Sosnowska} \& {Udry}}{{Pepe} et~al.}{2004}]{pepe04}
{Pepe} F.,  {Mayor} M.,  {Queloz} D.,  {Benz} W.,  {Bonfils} X.,  {Bouchy} F.,
  {Curto} G.~L.,  {Lovis} C.,  {M{\'e}gevand} D.,  {Moutou} C.,  {Naef} D.,
  {Rupprecht} G.,  {Santos} N.~C.,  {Sivan} J.-P.,  {Sosnowska} D.,    {Udry}
  S.,  2004, A\&A, 423, 385

\bibitem[\protect\citeauthoryear{Perrier, Sivian, Naef, Beuzit, Mayor, Queloz
  \& Udry}{Perrier et~al.}{2003}]{perrier03}
Perrier C.,  Sivian J.~P.,  Naef D.,  Beuzit J.~L.,  Mayor M.,  Queloz D.,
  Udry S.,  2003, A\&A, 410, 1039

\bibitem[\protect\citeauthoryear{Pizzolato, Maggio, Micela, Sciortino \&
  Ventura}{Pizzolato et~al.}{2003}]{pizzolato03}
Pizzolato N.,  Maggio A.,  Micela G.,  Sciortino S.,    Ventura P.,  2003,
  A\&A, 397, 147

\bibitem[\protect\citeauthoryear{{Pont}, {H\'{e}brard}, {Irwin}, {Bouchy},
  {Moutou}, {Ehrenreich}, {Guillot}, {Aigrain}, {Bonfils}, {Berta}, {Boisse},
  {Burke}, {Charbonneau}, {Delfosse}, {Desort}, {Eggenberger}, {Forveille} \&
  {et al.}}{{Pont} et~al.}{2009}]{pont09}
{Pont} F.,  {H\'{e}brard} G.,  {Irwin} J.~M.,  {Bouchy} F.,  {Moutou} C.,
  {Ehrenreich} D.,  {Guillot} T.,  {Aigrain} S.,  {Bonfils} X.,  {Berta} Z.,
  {Boisse} I.,  {Burke} C.,  {Charbonneau} D.,  {Delfosse} X.,  {Desort} M.,
  {Eggenberger} A.,  {Forveille} T.,    {et al.} 2009, A\&A, 502, 695

\bibitem[\protect\citeauthoryear{{Pont}, {Moutou}, {Gillon}, {Udalski},
  {Bouchy}, {Fernandes}, {Gieren}, {Mayor}, {Mazeh}, {Minniti}, {Melo}, {Naef},
  {Pietrzynski}, {Queloz}, {Ruiz}, {Santos} \& {Udry}}{{Pont}
  et~al.}{2007}]{pont07}
{Pont} F.,  {Moutou} C.,  {Gillon} M.,  {Udalski} A.,  {Bouchy} F.,
  {Fernandes} J.~M.,  {Gieren} W.,  {Mayor} M.,  {Mazeh} T.,  {Minniti} D.,
  {Melo} C.,  {Naef} D.,  {Pietrzynski} G.,  {Queloz} D.,  {Ruiz} M.~T.,
  {Santos} N.~C.,    {Udry} S.,  2007, A\&A, 465, 1069

\bibitem[\protect\citeauthoryear{Queloz, Anderson, Collier~Cameron, Gillon,
  Hebb, Hellier, Maxted, Pepe, Pollacco, S\'{e}gransan, Smalley, Triaud, Udry
  \& West}{Queloz et~al.}{2010}]{queloz10}
Queloz D.,  Anderson D.,  Collier~Cameron A.,  Gillon M.,  Hebb L.,  Hellier
  C.,  Maxted P.,  Pepe F.,  Pollacco D.,  S\'{e}gransan d.,  Smalley B.,
  Triaud A. H. M.~J.,  Udry S.,    West R.,  2010, A\&A, submitted

\bibitem[\protect\citeauthoryear{{Queloz}, {Eggenberger}, {Mayor}, {Perrier},
  {Beuzit}, {Naef}, {Sivan} \& {Udry}}{{Queloz} et~al.}{2000}]{queloz00}
{Queloz} D.,  {Eggenberger} A.,  {Mayor} M.,  {Perrier} C.,  {Beuzit} J.~L.,
  {Naef} D.,  {Sivan} J.~P.,    {Udry} S.,  2000, A\&A, 359, L13

\bibitem[\protect\citeauthoryear{Reiners \& Schmitt}{Reiners \&
  Schmitt}{2003}]{reiners03}
Reiners A.,  Schmitt J. H. M.~M.,  2003, A\&A, 398, 647

\bibitem[\protect\citeauthoryear{Ribas \& Miralda-Escud\'{e}}{Ribas \&
  Miralda-Escud\'{e}}{2007}]{ribas07}
Ribas I.,  Miralda-Escud\'{e} J.,  2007, A\&A, 464, 779

\bibitem[\protect\citeauthoryear{Saar \& Osten}{Saar \& Osten}{1997}]{saar97}
Saar S.~H.,  Osten R.~A.,  1997, MNRAS, 284, 803

\bibitem[\protect\citeauthoryear{Saffe, G\'{o}mez \& Chavero}{Saffe
  et~al.}{2005}]{saffe05}
Saffe C.,  G\'{o}mez M.,    Chavero C.,  2005, A\&A, 443, 609

\bibitem[\protect\citeauthoryear{{Santos}, {Bouchy}, {Mayor}, {Pepe}, {Queloz},
  {Udry}, {Lovis}, {Bazot}, {Benz}, {Bertaux}, {Lo Curto}, {Delfosse},
  {Mordasini}, {Naef}, {Sivan} \& {Vauclair}}{{Santos} et~al.}{2004}]{santos04}
{Santos} N.~C.,  {Bouchy} F.,  {Mayor} M.,  {Pepe} F.,  {Queloz} D.,  {Udry}
  S.,  {Lovis} C.,  {Bazot} M.,  {Benz} W.,  {Bertaux} J.-L.,  {Lo Curto} G.,
  {Delfosse} X.,  {Mordasini} C.,  {Naef} D.,  {Sivan} J.-P.,    {Vauclair} S.,
   2004, A\&A, p.~L19

\bibitem[\protect\citeauthoryear{{Santos}, {Mayor}, {Naef}, {Pepe}, {Queloz},
  {Udry} \& {Blecha}}{{Santos} et~al.}{2000}]{santos00}
{Santos} N.~C.,  {Mayor} M.,  {Naef} D.,  {Pepe} F.,  {Queloz} D.,  {Udry} S.,
    {Blecha} A.,  2000, A\&A, 361, 265

\bibitem[\protect\citeauthoryear{{Santos}, {Mayor}, {Naef}, {Pepe}, {Queloz},
  {Udry}, {Burnet}, {Clausen}, {Helt}, {Olsen} \& {Pritchard}}{{Santos}
  et~al.}{2002}]{santos02}
{Santos} N.~C.,  {Mayor} M.,  {Naef} D.,  {Pepe} F.,  {Queloz} D.,  {Udry} S.,
  {Burnet} M.,  {Clausen} J.~V.,  {Helt} B.~E.,  {Olsen} E.~H.,    {Pritchard}
  J.~D.,  2002, A\&A, 392, 215

\bibitem[\protect\citeauthoryear{{Santos}, {Mayor}, {Naef} D.and~{Pepe},
  {Queloz}, {Udry} \& {Burnet}}{{Santos} et~al.}{2001}]{santos01}
{Santos} N.~C.,  {Mayor} M.,  {Naef} D.and~{Pepe} F.,  {Queloz} D.,  {Udry} S.,
     {Burnet} M.,  2001, A\&A, 379, 999

\bibitem[\protect\citeauthoryear{{Santos}, {Pont}, {Melo}, {Israelian},
  {Bouchy}, {Mayor}, {Moutou}, {Queloz}, {Udry} \& {Guillot}}{{Santos}
  et~al.}{2006}]{santos06}
{Santos} N.~C.,  {Pont} F.,  {Melo} C.,  {Israelian} G.,  {Bouchy} F.,  {Mayor}
  M.,  {Moutou} C.,  {Queloz} D.,  {Udry} S.,    {Guillot} T.,  2006, A\&A,
  450, 825

\bibitem[\protect\citeauthoryear{{Sato}, {Ando}, {Kambe}, {Takeda}, {Izumiura},
  {Masuda}, {Watanabe}, {Noguchi}, {Wada}, {Okada}, {Koyano}, {Maehara},
  {Norimoto}, {Okada}, {Shimizu}, {Uraguchi}, {Yanagisawa} \& {Yoshida}}{{Sato}
  et~al.}{2003}]{sato03}
{Sato} B.,  {Ando} H.,  {Kambe} E.,  {Takeda} Y.,  {Izumiura} H.,  {Masuda} S.,
   {Watanabe} E.,  {Noguchi} K.,  {Wada} S.,  {Okada} N.,  {Koyano} H.,
  {Maehara} H.,  {Norimoto} Y.,  {Okada} T.,  {Shimizu} Y.,  {Uraguchi} F.,
  {Yanagisawa} K.,    {Yoshida} M.,  2003, ApJ, 597, L157

\bibitem[\protect\citeauthoryear{{Simpson}, {Pollacco}, {H{\'e}brard},
  {Gibson}, {Barros}, {Boisse}, {Bouchy}, {Cameron}, {Miller}, {Watson} \&
  {Keenan}}{{Simpson} et~al.}{2010}]{simpson10}
{Simpson} E.~K.,  {Pollacco} D.,  {H{\'e}brard} G.,  {Gibson} N.~P.,  {Barros}
  S.~C.~C.,  {Boisse} I.,  {Bouchy} F.,  {Cameron} A.~C.,  {Miller} G.~R.~M.,
  {Watson} C.~A.,    {Keenan} F.~P.,  2010, MNRAS, accepted

\bibitem[\protect\citeauthoryear{Skelly, Unruh, Barnes, Lawson, Donati \&
  Collier~Cameron}{Skelly et~al.}{2009}]{skelly09}
Skelly M.~B.,  Unruh Y.~C.,  Barnes J.~R.,  Lawson W.~A.,  Donati J.-F.,
  Collier~Cameron A.,  2009, MNRAS, 399, 1829

\bibitem[\protect\citeauthoryear{{Soderblom}}{{Soderblom}}{1982}]{soderblom82}
{Soderblom} D.~R.,  1982, ApJ, 263, 239

\bibitem[\protect\citeauthoryear{{Sozzetti}, {Torres}, {Charbonneau}, {Latham},
  {Holman}, {Winn}, {Laird} \& {O'Donovan}}{{Sozzetti}
  et~al.}{2007}]{sozzetti07}
{Sozzetti} A.,  {Torres} G.,  {Charbonneau} D.,  {Latham} D.~W.,  {Holman}
  M.~J.,  {Winn} J.~N.,  {Laird} J.~B.,    {O'Donovan} F.~T.,  2007, ApJ, 664,
  1190

\bibitem[\protect\citeauthoryear{{Sozzetti}, {Udry}, {Zucker}, {Torres},
  {Beuzit}, {Latham}, {Mayor}, {Mazeh}, {Naef}, {Perrier}, {Queloz} \&
  {Sivan}}{{Sozzetti} et~al.}{2006}]{sozzetti06}
{Sozzetti} A.,  {Udry} S.,  {Zucker} S.,  {Torres} G.,  {Beuzit} J.~L.,
  {Latham} D.~W.,  {Mayor} M.,  {Mazeh} T.,  {Naef} D.,  {Perrier} C.,
  {Queloz} D.,    {Sivan} J.-P.,  2006, A\&A, 449, 417

\bibitem[\protect\citeauthoryear{{Sozzetti}, {Yong}, {Torres}, {Charbonneau},
  {Latham}, {Allende Prieto}, {Brown}, {Carney} \& {Laird}}{{Sozzetti}
  et~al.}{2004}]{sozzetti04}
{Sozzetti} A.,  {Yong} D.,  {Torres} G.,  {Charbonneau} D.,  {Latham} D.~W.,
  {Allende Prieto} C.,  {Brown} T.~M.,  {Carney} B.~W.,    {Laird} J.~B.,
  2004, ApJ, 616, L167

\bibitem[\protect\citeauthoryear{Strassmeier, Washuetti, Granzer, Scheck \&
  Weber}{Strassmeier et~al.}{2000}]{strassmeier00b}
Strassmeier K.~G.,  Washuetti A.,  Granzer T.,  Scheck M.,    Weber M.,  2000,
  A\&AS, 142, 275

\bibitem[\protect\citeauthoryear{Takeda \& Rasio}{Takeda \&
  Rasio}{2005}]{takeda05}
Takeda G.,  Rasio F.~A.,  2005, ApJ, 627, 1001

\bibitem[\protect\citeauthoryear{{Tegmark}, {Strauss}, {Blanton}, {Abazajian},
  {Dodelson}, {Sandvik}, {Wang}, {Weinberg}, {Zehavi}, {Bahcall}, {Hoyle},
  {Schlegel}, {Scoccimarro}, {Vogeley}, {Berlind}, {Budavari} \& {et
  al.}}{{Tegmark} et~al.}{2004}]{tegmark04}
{Tegmark} M.,  {Strauss} M.~A.,  {Blanton} M.~R.,  {Abazajian} K.,  {Dodelson}
  S.,  {Sandvik} H.,  {Wang} X.,  {Weinberg} D.~H.,  {Zehavi} I.,  {Bahcall}
  N.~A.,  {Hoyle} F.,  {Schlegel} D.,  {Scoccimarro} R.,  {Vogeley} M.~S.,
  {Berlind} A.,  {Budavari} T.,    {et al.} 2004, Phys. Rev. D., 69, 103501

\bibitem[\protect\citeauthoryear{Tingley \& Sackett}{Tingley \&
  Sackett}{2005}]{tingley05}
Tingley B.,  Sackett P.~D.,  2005, AJ, 627, 1011

\bibitem[\protect\citeauthoryear{{Torres}, {Winn} \& {Holman}}{{Torres}
  et~al.}{2008}]{torres08}
{Torres} G.,  {Winn} J.~N.,    {Holman} M.~J.,  2008, ApJ, 677, 1324

\bibitem[\protect\citeauthoryear{Triaud, Collier~Cameron, Queloz, Anderson,
  Gillon, Hebb, Hellier, Loeillet, Maxted, Mayor, Pepe, Pollacco,
  S\'{e}gransan, Smalley, Udry, West \& Wheatley}{Triaud
  et~al.}{2010}]{triaud10}
Triaud A. H. M.~J.,  Collier~Cameron A.,  Queloz D.,  Anderson D.~R.,  Gillon
  M.,  Hebb L.,  Hellier C.,  Loeillet B.,  Maxted P. F.~L.,  Mayor M.,  Pepe
  F.,  Pollacco D.,  S\'{e}gransan D.,  Smalley B.,  Udry S.,  West R.~G.,
  Wheatley P.~J.,  2010, A\&A, submitted

\bibitem[\protect\citeauthoryear{{Triaud}, {Queloz}, {Bouchy}, {Moutou},
  {Cameron}, {Claret}, {Barge}, {Benz}, {Deleuil}, {Guillot}, {H{\'e}brard},
  {Lecavelier Des {\'E}tangs}, {Lovis}, {Mayor}, {Pepe} \& {Udry}}{{Triaud}
  et~al.}{2009}]{triaud09}
{Triaud} A.~H.~M.~J.,  {Queloz} D.,  {Bouchy} F.,  {Moutou} C.,  {Cameron}
  A.~C.,  {Claret} A.,  {Barge} P.,  {Benz} W.,  {Deleuil} M.,  {Guillot} T.,
  {H{\'e}brard} G.,  {Lecavelier Des {\'E}tangs} A.,  {Lovis} C.,  {Mayor} M.,
  {Pepe} F.,    {Udry} S.,  2009, A\&A, 506, 377

\bibitem[\protect\citeauthoryear{Udry, Mayor, Benz, Bertaux, Bouchy, Lovis,
  Mordasini, Pepe, Queloz \& Sivan}{Udry et~al.}{2006}]{udry06}
Udry S.,  Mayor M.,  Benz W.,  Bertaux J.~L.,  Bouchy F.,  Lovis C.,  Mordasini
  C.,  Pepe F.,  Queloz D.,    Sivan J.~P.,  2006, A\&A, 447, 361

\bibitem[\protect\citeauthoryear{{Udry}, {Mayor}, {Clausen}, {Freyhammer},
  {Helt}, {Lovis}, {Naef}, {Olsen}, {Pepe}, {Queloz} \& {Santos}}{{Udry}
  et~al.}{2003}]{udry03}
{Udry} S.,  {Mayor} M.,  {Clausen} J.~V.,  {Freyhammer} L.~M.,  {Helt} B.~E.,
  {Lovis} C.,  {Naef} D.,  {Olsen} E.~H.,  {Pepe} F.,  {Queloz} D.,    {Santos}
  N.~C.,  2003, A\&A, 407, 679

\bibitem[\protect\citeauthoryear{Udry, Mayor, Naef, Pepe, Queloz, Santos \&
  Burnet}{Udry et~al.}{2002}]{udry02}
Udry S.,  Mayor M.,  Naef D.,  Pepe F.,  Queloz D.,  Santos N.~C.,    Burnet
  M.,  2002, A\&A, 390, 267

\bibitem[\protect\citeauthoryear{Valenti \& Fischer}{Valenti \&
  Fischer}{2005}]{valenti05}
Valenti J.~A.,  Fischer D.~A.,  2005, ApJS, 159, 141

\bibitem[\protect\citeauthoryear{Vaughan, Baliunas, Middelkoop, Hartmann,
  Mihalas, Noyes \& Preston}{Vaughan et~al.}{1981}]{vaughan81}
Vaughan A.~H.,  Baliunas S.~L.,  Middelkoop F.,  Hartmann L.~W.,  Mihalas D.,
  Noyes R.~W.,    Preston G.~W.,  1981, ApJ, 250, 276

\bibitem[\protect\citeauthoryear{{Vogt}, {Butler}, {Marcy}, {Fischer},
  {Pourbaix}, {Apps} \& {Laughlin}}{{Vogt} et~al.}{2002}]{vogt02}
{Vogt} S.~S.,  {Butler} R.~P.,  {Marcy} G.~W.,  {Fischer} D.~A.,  {Pourbaix}
  D.,  {Apps} K.,    {Laughlin} G.,  2002, ApJ, 568, 352

\bibitem[\protect\citeauthoryear{Watson, Dhillon \& Shabaz}{Watson
  et~al.}{2006}]{watson06}
Watson C.~A.,  Dhillon V.~S.,    Shabaz T.,  2006, MNRAS, 368, 637

\bibitem[\protect\citeauthoryear{Watson, Steeghs, Shabaz \& Dhillon}{Watson
  et~al.}{2007}]{watson07}
Watson C.~A.,  Steeghs D.,  Shabaz T.,    Dhillon V.~S.,  2007, MNRAS, 382,
  1105

\bibitem[\protect\citeauthoryear{Wilson}{Wilson}{1978}]{wilson78}
Wilson O.~C.,  1978, ApJ, 226, 379

\bibitem[\protect\citeauthoryear{{Winn}, {Howard}, {Johnson}, {Marcy}, {Gazak},
  {Starkey}, {Ford}, {Col{\'o}n}, {Reyes}, {Nortmann}, {Dreizler}, {Odewahn},
  {Welsh}, {Kadakia}, {Vanderbei}, {Adams}, {Lockhart}, {Crossfield} \& et
  al.}{{Winn} et~al.}{2009}]{winn09}
{Winn} J.~N.,  {Howard} A.~W.,  {Johnson} J.~A.,  {Marcy} G.~W.,  {Gazak}
  J.~Z.,  {Starkey} D.,  {Ford} E.~B.,  {Col{\'o}n} K.~D.,  {Reyes} F.,
  {Nortmann} L.,  {Dreizler} S.,  {Odewahn} S.,  {Welsh} W.~F.,  {Kadakia} S.,
  {Vanderbei} R.~J.,  {Adams} E.~R.,  {Lockhart} M.,  {Crossfield} I.~J.,    et
  al. 2009, ApJ, 703, 2091

\bibitem[\protect\citeauthoryear{{Winn}, {Johnson}, {Narita}, {Suto}, {Turner},
  {Fischer}, {Butler}, {Vogt}, {O'Donovan} \& {Gaudi}}{{Winn}
  et~al.}{2008}]{winn08}
{Winn} J.~N.,  {Johnson} J.~A.,  {Narita} N.,  {Suto} Y.,  {Turner} E.~L.,
  {Fischer} D.~A.,  {Butler} R.~P.,  {Vogt} S.~S.,  {O'Donovan} F.~T.,
  {Gaudi} B.~S.,  2008, ApJ, 682, 1283

\bibitem[\protect\citeauthoryear{{Winn}, {Johnson}, {Peek}, {Marcy}, {Bakos},
  {Enya}, {Narita}, {Suto}, {Turner} \& {Vogt}}{{Winn} et~al.}{2007}]{winn07}
{Winn} J.~N.,  {Johnson} J.~A.,  {Peek} K.~M.~G.,  {Marcy} G.~W.,  {Bakos}
  G.~{\'A}.,  {Enya} K.,  {Narita} N.,  {Suto} Y.,  {Turner} E.~L.,    {Vogt}
  S.~S.,  2007, ApJ, 665, L167

\bibitem[\protect\citeauthoryear{Winn, Noyes, Holman, {Charbonneau}, {Ohta},
  {Taruya}, {Suto}, {Narita}, {Turner}, {Johnson}, {Marcy}, {Butler} \&
  {Vogt}}{Winn et~al.}{2005}]{winn05}
Winn J.~N.,  Noyes R.~W.,  Holman M.~J.,  {Charbonneau} D.,  {Ohta} Y.,
  {Taruya} A.,  {Suto} Y.,  {Narita} N.,  {Turner} E.~L.,  {Johnson} J.~A.,
  {Marcy} G.~W.,  {Butler} R.~P.,    {Vogt} S.~S.,  2005, ApJ, 631, 1215

\bibitem[\protect\citeauthoryear{{Wolf}, {Laughlin}, {Henry}, {Fischer},
  {Marcy}, {Butler} \& {Vogt}}{{Wolf} et~al.}{2007}]{wolf07}
{Wolf} A.~S.,  {Laughlin} G.,  {Henry} G.~W.,  {Fischer} D.~A.,  {Marcy} G.,
  {Butler} P.,    {Vogt} S.,  2007, ApJ, 667, 549

\bibitem[\protect\citeauthoryear{Wolszcan \& Frail}{Wolszcan \&
  Frail}{1992}]{wolszcan92}
Wolszcan A.,  Frail D.~A.,  1992, Nature, 355, 145

\bibitem[\protect\citeauthoryear{Wright, Marcy, Butler \& Vogt}{Wright
  et~al.}{2004}]{wright04}
Wright J.~T.,  Marcy G.~W.,  Butler P.~R.,    Vogt S.~S.,  2004, ApJS, 152, 261

\end{thebibliography}

\end{document}